\documentclass[12pt]{article}
\usepackage{epsfig,amsfonts,amssymb,amsmath}
\usepackage{hyperref}
\usepackage{cite,upgreek}
\input epsf.sty
\usepackage{array}
\newcolumntype{P}[1]{>{\centering\arraybackslash}p{#1}}
\newcolumntype{M}[1]{>{\centering\arraybackslash}m{#1}}

\def\ZZZ{{\hbox{ Z\kern-1.6mm Z}}}
\def\RRR{{\hbox{ R\kern-2.4mm R}}}
\def\CCC{{\hbox{ C\kern-2.0mm C}}}
\def\zzz{{\hbox{z\kern-1mm z}}}

\newcommand{\f}{\frac}

\newcommand{\qeq}{{\hbox{=\kern-2.3mm ? \kern.5mm }}}
\renewcommand{\qeq}{=}

\newcommand{\non}{\nonumber}
\newcommand{\be}{\begin{eqnarray}}
\newcommand{\ee}{\end{eqnarray}}
\newcommand{\ben}{\begin{eqnarray}\displaystyle}
\newcommand{\een}{\end{eqnarray}}

\newcommand{\p}{\partial}
\newcommand{\sectiono}[1]{\section{#1}\setcounter{equation}{0}}
\newcommand{\subsectiono}[1]{\subsection{#1}\setcounter{equation}{0}}

\def\one{{\hbox{ 1\kern-.8mm l}}}
\def\zero{{\hbox{ 0\kern-1.5mm 0}}}

\newcommand{\bea}[1]{\begin{eqnarray}\label{#1} }
\newcommand{\eea}{\end{eqnarray}}

\usepackage{bm}
\usepackage[table]{xcolor}

\begin{document}

\begin{flushright}
HRI/ST/1704 
\end{flushright}

\baselineskip 24pt

\begin{center}
{\Large \bf  Amplitudes Involving Massive States Using Pure Spinor Formalism}

\end{center}

\vskip .6cm
\medskip

\vspace*{4.0ex}

\baselineskip=18pt

\begin{center}

{\large 
\rm Subhroneel Chakrabarti$^{a}$, Sitender Pratap Kashyap$^{a}$, Mritunjay Verma$^{a,b}$ }

\end{center}

\vspace*{4.0ex}

\centerline{\it \small $^a$Harish-Chandra Research Institute, HBNI, 
Chhatnag Road, Jhunsi,
Allahabad 211019, India}
\centerline{ \it \small $^b$International Centre for Theoretical Sciences,
 Hesarghatta,
Bengaluru 560089, India.}

\vspace*{1.0ex}
\centerline{\small E-mail: subhroneelchak,
sitenderpratap, mritunjayverma@hri.res.in}

\vspace*{5.0ex}

\centerline{\bf Abstract} \bigskip
Same amplitudes evaluated independently using RNS and pure spinor formalism are expected to agree. While for massless states, this fact has been firmly established, for massive states such an explicit check has been lacking so far. We compute all massless-massless-massive 3-point functions in open supertrings in pure spinor formalism for the first massive states and compare them with the corresponding RNS results. We fix the normalization of the vertex operators of the massive states by comparing same set of 3-point functions for a fixed ordering \color{black} in the two formalisms. Once fixed, the subsequent 3-point functions for each inequivalent ordering match exactly\color{black}. This extends the explicit demonstration of equivalence of pure spinor and RNS formalism from massless states to first massive states.

\vfill

%\begin{center}
%{\Huge Preliminary version}
%\end{center}

\vfill \eject

\baselineskip 18pt

\tableofcontents

\sectiono{Introduction } 
Since the turn of last century, the pure spinor formalism has emerged as a powerful alternative to the RNS and Green Schwarz formalisms for superstring theory \cite{Berkovits, Berkovits7, BerkovitsNekrasov}. The main advantage of the pure spinor formalism over RNS is that it maintains manifest spacetime super-Poincar\'e covariance at all stages. There are several general arguments demonstrating how the cohomology of the pure spinor formalism is equivalent to the cohomology of the RNS and Green Schwarz formalisms \cite{Berkovits3, Berkovits4, Berkovits8}. For massless states several explicit calculations of scattering amplitude in pure spinor formalism have allowed a direct comparison with the corresponding amplitude calculation in RNS formalism \cite{Berkovits9, Berkovits_Mafra, Berkovits10, Gomez:2010ad}. This direct comparison fixed the relative normalization between the vertex operators in PS and RNS formalism for the massless states.

\vspace*{.07in} In this paper we extend such explicit direct comparison for the first massive states ($m^2 =\frac{1}{\alpha'}$) of open superstring. We explicitly compute all possible tree level 3 point functions for a fixed ordering of vertex operators \color{black} in PS formalism involving 2 massless and 1 massive state and compare them directly with their RNS counterparts. The massless spectrum contains a bosonic gluon field $a_m$ (with 8 degrees of freedom) and a fermionic gluino field $\chi^\alpha$(with 8 degrees of freedom). The first massive states include a bosonic 3 form field $b_{mnp}$ (with 84 degrees of freedom), a bosonic symmetric rank 2 tensor field $g_{mn}$ (with 44 degrees of freedom) and a fermionic spin $\frac{3}{2}$ field $\psi_{m\alpha}$ (with 128 = 84+44 degrees of freedom).

\vspace*{.07in}The computation of the above mentioned 3 point functions in PS formalism requires the massless unintegrated vertex and its covariant $\theta$ expansion \cite{Berkovits, Harnad, Ooguri, Policastro} along with the massive unintegrated vertex and its covariant $\theta$ expansion \cite{Berkovits1, theta_exp}. While the tree level amplitude prescription in PS is straightforward, one typically encounters a large number of terms in $\theta$ expansion of massive vertex and the calculation becomes cumbersome to compute solely by hand. For this reason, the computation was done using Cadabra \cite{0608005,0701238} and benchmarked by computing 3-point massless-massless-massless amplitudes which are already known in literature (see, e.g. \cite{mafrathesis}). For use of Cadabra in pure spinor calculations, also see \cite{Suna:2016bzw}. The final amplitudes were computed directly using this code and compared with their corresponding RNS results. After fixing the normalization of the three massive vertex operators by comparing three correlators (for three massive states), we find that the rest of the correlators agree perfectly.

\vspace*{.07in}The rest of the paper is organized as follows. In section \ref{sec:review}, we review the tree level amplitude prescription in PS. In section \ref{thetaex}, we summarize the $\theta$ expansion procedure for massive states given in \cite{ theta_exp} and give the result for the $\theta$ expansion to the relevant order we need in this work. Section \ref{sec:3pt_PS} gives the detailed derivation and results of 3-point amplitudes in PS formalism. In section \ref{sec:compare}, we compare our results obtained in PS formalism directly with the amplitudes computed using RNS formalism. We conclude in section \ref{outlook} with some brief discussion. In appendix \ref{appen:PS}, we give our conventions for the PS formalism. In appendix \ref{appen:RNS_results}, we give our conventions for the RNS calculations and summarize the results for 3-point functions in the RNS formalism. Finally, in appendix \ref{appen:massless_theta}, we give the theta expansion of the massless vertex operator in our conventions.

 \section{Tree amplitude prescription in pure spinor formalism}
% \section{Brief review of some pure spinor elements}
\label{sec:review}
Both minimal as well as the non minimal pure spinor formalisms give the same amplitude prescription at tree level. In this section, we shall review this prescription \cite{Berkovits, Berkovits7}. In particular, we shall focus on 3-point functions on the disk  with a specific ordering of vertex operators on the boundary. To evaluate any 3-point correlator, the knowledge of the unintegrated form of the vertex operator for external states is sufficient. All the amplitudes of interest in this paper are of the form 
\be
\mathcal{A}_3&=&\langle V_1V_2V_3 \rangle\ 
\ee
where $\langle V_1V_2V_3 \rangle $ denotes 3-point function on the disk with a fixed ordering.

\vspace*{.07in}In this paper,  $V_1$ and $V_2$ will be taken to be unintegrated vertex operators creating massless states (gluon or gluino) and $V_3$ will be the unintegrated vertex operator creating a massive state ($b_{mnp}, g_{mn}$ or $\psi_{m\alpha}$). However, we must emphasize, none of the strategy that we shall outline below is particular to this specific kind of 3-point functions. The tree-level amplitude prescription will be equally valid for any 3-point functions of open strings states (massive or massless).

Our choice of normalization of pure spinor measure is the standard one in the literature
\be \label{PS_measure} 
\left\langle (\lambda \gamma^s \theta)(\lambda \gamma^t \theta)(\lambda \gamma^u \theta)(\theta \gamma_{stu} \theta) \right\rangle _{PSS} = 1
\ee
The process of 3-point amplitude computation can be succinctly summarized in a series of steps that is given below.
\begin{itemize}
	\item{\bf Step 1 :} Assume a particular order of the vertex operator inserted on the disk and make use of the OPEs between various conformal weight 1 objects to reduce the three point function to a correlation function involving only the three superfields and three pure spinor ghost $\lambda^\alpha$ coming from the three vertices. The correlation function which contains $\p \theta^\alpha$ terms do not contribute since they always fail to have the correct number of $\theta$ zero modes to give non-zero answer as required by \ref{PS_measure}. \color{black} Now perform the $\theta$ expansion for each of the vertex, viz. $V_1$, $V_2$ and $V_3$, explicitly. Although one can do the full $\theta$ expansion, but that is rendered redundant due to \ref{PS_measure} which states that only terms involving exactly five $\theta$s can give a non-zero contribution. The relevant order of $\theta$ expansion for each of the vertex for a given amplitude can therefore be deduced from this consideration. 
	
	\item{\bf Step 2 :} From the product $V_1 V_2 V_3$, retain only the terms which have precisely five $\theta$s as all other terms will give zero contribution trivially due to \ref{PS_measure}. \footnote{Notice that \ref{PS_measure} also suggests we should retain only terms with exactly three number of $\lambda$s. But this is automatically ensured by the fact that by construction all unintegrated vertex in PS formalism have ghost number 1 and therefore they each carry exactly a single factor of $\lambda$. Consequentially the product $V_1 V_2 V_3$ always goes as $\lambda^3$. }
	
	\item{\bf Step 3 :} Expand the physical fields appearing in various vertices in a basis of plane waves with polarizations as the coefficients
	\be 
	H_{p_1...p_n}^{\beta_1...\beta_k}(X)= h _{p_1...p_n}^{\beta_1...\beta_k} \, e^{ik\cdot X}\non
	\ee
	Here $ h _{p_1...p_n}^{\beta_1...\beta_k}$ are the constant tensor-spinor of appropriate index structure denoting the polarizations for the physical field $H_{p_1...p_n}^{\beta_1...\beta_k}(X)$.
	
	The correlation function  $\langle V_1 V_2 V_3 \rangle $ at this stage factorizes completely into two separate correlation functions, one involving $X^m$ fields and the other involving pure spinor fields. These can be evaluated independently and their result can be multiplied to obtain the final answer for a given ordering.
	
	\item{\bf step 4 :} Evaluate the 3-point function on disk by the usual methodology of bosonic open strings. Typically these correlation functions can be put into the schematic form $\langle :e^{i k_1 \cdot X(x_1)}: :e^{i k_2 \cdot X(x_2)}: :\mathcal{F}(\p X^m(x_3)) \, e^{i k_3 \cdot X(x_3)}: \rangle_{Disk}$ where  $\mathcal{F} (\p X^m(x)) = \prod\limits_{i} \p X^{m_i}(x)$.
	
	\item{\bf Step 5 :} The correlators living on pure spinor superspace can be evaluated following the list of identities first derived in \cite{ Berkovits_Mafra, 0704.0015}. We list in appendix \ref{PureSuperspace} the subset of those identities that were needed in evaluating the 3-point functions in our case.
	
	\item {\bf Step 6 :} Multiply the results obtained from step 4 and step 5 to obtain the full contribution for a given order in theta expansion of the 3 vertices.

\end{itemize}

 Once the answer has been obtained for a given order, the final answer for all other inequivalent orders can be readily obtained by suitable permutation of momenta and polarization labels. Since each ordering carries different Chan-Paton factors, the full amplitude cannot be obtained by simply adding the contribution coming from different ordering before multiplying them by Chan-Paton factors. Therefore, to compare our result with the result obtained in RNS formalism, we shall directly compare each inequivalent ordering separately.

\vspace*{.06in}While the algorithm described above is quite straightforward in principle, the growing number of terms in $\theta$ expansion of the vertex operators implies that the cleanest way to perform these amplitude calculations beyond a point is to employ the help of an available computer algebra system. As mentioned in the introduction, we used Cadabra \cite{0608005,0701238} which is an open source computer algebra system developed to aid in field-theoretic computations. For our present purpose, we found Cadabra to be unparalleled in implementing the algorithm described above.

\section{First massive vertex operator and its $\theta$  expansion} \label{thetaex}

At the 1st mass level of the open string, there are 128 fermionic and 128 bosonic degrees of freedom. The fermionic degrees of freedom are contained in a spin-$3/2$ field $\psi_{m\alpha}$ satisfying
\be 
\p^m\psi_{m\alpha}=0 \qquad ; \qquad \gamma^{m\alpha\beta} \psi_{m\beta}=0  .  \label{constraints_theta_ind_comp}
\ee
The bosonic degrees of freedom are contained in a traceless symmetric tensor $g_{mn}$ and a 3-form field $b_{mnp}$ satisfying
\be 
\p^m b_{mnp}=0 \qquad ;\qquad
\eta^{mn}g_{mn}=0\qquad ;\qquad  \p^m g_{mn}=0  . \label{constraints_theta_ind_comp1}
\ee
The constraints \eqref{constraints_theta_ind_comp} and \eqref{constraints_theta_ind_comp1} ensure that the number of independent components in the fields $\psi_{m\beta}, b_{mnp}$ and $g_{mn}$ are $128, 84$ and $44$ respectively. These fields form a massive spin-2 supermultiplet in 10 dimensions. In pure spinor formalism, we use the language of superspace to describe the system in a manifestly supersymmetric invariant manner. This is done by introducing three basic superfields $\Psi_{m\alpha}, B_{mnp}$ and $G_{mn}$ whose theta independent components are $\psi_{m\alpha}, b_{mnp}$ and $g_{mn}$ respectively. These basic superfields satisfy the superspace equations \cite{theta_exp}
 \be
D_{\alpha}G_{sm} = 16\ \p^{p}(\gamma_{p(s}\Psi_{m)})_\alpha \label{D_Gmn1}
\ee
\be
D_\alpha B_{mnp}=12 (\gamma_{[mn}\Psi_{p]})_\alpha - 24\alpha' \p^t \p_{[m}(\gamma_{|t|n}\Psi_{p]})_\alpha                             \label{D_Bmnp_general}
\ee
\be
D_{\alpha}\Psi_{s\beta}= \f{1}{16} G_{sm}\gamma^{m}_{\alpha\beta}+\f{1}{24}\p_mB_{nps}(\gamma^{mnp})_{\alpha\beta}
-\f{1}{144}\p^mB^{npq}(\gamma_{smnpq})_{\alpha\beta} \label{D_Psi1}
\ee
and the constraints
\be
 (\gamma^{m})^{\alpha \beta}\Psi_{m\beta}=0\quad ; \quad \p^m \Psi_{m\beta}=0\quad ; \quad  \p^m B_{mnp}=0 \quad ; \quad  \p^m G_{m n}=0 \;\; \;; \;\; \eta^{mn}G_{mn}=0%\label{cons_theta=0}
\ee
where, $D_\alpha=\p_\alpha +\gamma^m_{\alpha\beta}\theta^\beta\p_m$.

\vspace*{.07in}The equations \eqref{D_Gmn1}, \eqref{D_Bmnp_general} and \eqref{D_Psi1} give recursion relations to determine the complete theta expansion of the superfields $B_{mnp}, G_{mn}$ and $\Psi_{m\alpha}$ \cite{theta_exp}. Before giving the theta expansion results, we note the unintegrated vertex operator for the 1st massive states of the open string which can be expressed in terms of the basic superfields as \cite{Berkovits1}
\ben
V\ =\ \partial \theta^\beta \lambda^\alpha B_{\alpha\beta} 
\ +\ d_\beta\lambda^\alpha C^\beta_{\;\alpha}\ +\ \Pi^m\lambda^\alpha H_{m\alpha}
\ +\ N^{mn}\lambda^\alpha F_{\alpha mn}                                                    \label{vertex_ansatz}
\een
where \cite{Berkovits1,theta_exp}, 
\be 
&&H_{m\alpha}=-72 \Psi_{m\alpha}\;, \quad C_{mnpq}=\f{1}{2}\p_{[m}B_{npq]} \;,  \non\\[.2in]
&&\hspace{.2 in}F_{\alpha mn} =-9\Bigl(7\p_{[m}\Psi_{n]\alpha} + \p^q(\gamma_{q[m})_\alpha^{\;\;\beta}\Psi_{n]\beta}\Bigl).
\label{2.14}
\ee
For calculating any amplitude involving the massive states in pure spinor formalism, we shall need the theta expansion result of the above vertex operator. Clearly, the theta expansion of the basic superfields $B_{mnp}$ and $\Psi_{m\alpha}$ automatically implies the theta expansion of the full vertex operator. For our purposes, we shall need the theta expansion results of $\Psi_{m\alpha} $ upto order $\theta^3$ and that of $B_{\alpha\beta}$ upto order $\theta^4$. We used the mathematica package GAMMA to do this computation\cite{Gamma}. Using equations \eqref{D_Gmn1}, \eqref{D_Bmnp_general} and \eqref{D_Psi1}, we obtain (for details, please see \cite{theta_exp})
\be
&&\Psi_{s\beta}\non\\
&=&\psi_{s\beta}+\f{1}{16}(\gamma^m\theta)_\beta\ g_{sm}-\f{i}{24}(\gamma^{mnp}\theta)_\beta k_{m}b_{nps}-\f{i}{144}(\gamma_{s}^{\;\;npqr}\theta)_{\beta}k_{n}b_{pqr} \non\\
&&-\f{i}{2}k^p(\gamma^m\theta)_{\beta}(\psi_{(m}\gamma_{s)p}\theta)-\f{i}{4}k_{m}(\gamma^{mnp}\theta)_{\beta}(\psi_{[s}\gamma_{np]}\theta)-\f{i}{24}(\gamma_{s}^{\;\;mnpq}\theta)_{\beta}k_{m}(\psi_{q}\gamma_{np}\theta)\non\\
&&-\f{i}{6}\alpha'k_mk^rk_s(\gamma^{mnp}\theta)_{\beta}
(\psi_p\gamma_{rn}\theta)+{\f{i}{288}}\alpha'(\gamma^{mnp}\theta)_\beta k_mk^rk_s(\theta\gamma^q_{\;\;nr}\theta)\ g_{pq}\non\\
&&-{\f{i}{192}}(\gamma^{mnp}\theta)_\beta k_m(\theta\gamma^q_{\;\;[np}\theta) g_{s]q}-\f{i}{1152}(\gamma_{smnpq}\theta)_{\beta}k^{m}(\theta\gamma_{npt}\theta)\ g^{qt}\non\\
&&-\f{i}{96} k^p(\gamma^m\theta)_{\beta}(\theta\gamma_{pq(s}\theta)\ g_{m)q}-\f{1}{1728}(\gamma^{mnp}\theta)_\beta k_m(\theta\gamma^{tuvw}_{\;\;\;\;\;\;\;\;\;nps}\theta)k_tb_{uvw}\non\\
&&-{\f{1}{864\alpha'}}(\gamma_s\theta)_{\beta}(\theta\gamma^{npq}\theta)b_{npq}-\f{1}{10368}(\gamma_{s}^{\;\;mnpq}\theta)_{\beta}k_{m}(\theta\gamma_{tuvwnpq}\theta)k^tb^{uvw}\non\\
&&-{\f{1}{864}}(\gamma^m\theta)_{\beta}(\theta\gamma^{npq}\theta)b_{npq}k_{m}k_s-\f{1}{576}(\gamma_{smnpq}\theta)_{\beta}k^{m}(\theta\gamma^{tun}\theta)b_u^{\;\;pq}k_t\non\\
&&-\f{1}{96\alpha'}(\gamma^m\theta)_{\beta}(\theta\gamma^{qr}_{\;\;\;\;(s}\theta)b_{m)rq}+\f{1}{96}(\gamma^m\theta)_{\beta}(\theta\gamma^{nqr}\theta)k_nk_{(s}b_{m)qr}\non\\
&&+{\f{1}{96}}(\gamma^{mnp}\theta)_\beta k_m(\theta\gamma^r_{\;\;q[n}\theta)b_{ps]r} k^{q} +O(\theta^4)
\ee
%}}
Similarly, the $\theta$ expansion of the superfield $B_{\alpha\beta}$ is given by
\be
B_{\alpha\beta}&=&\gamma^{mnp}_{\alpha\beta}\Biggl[b_{mnp}+12(\psi_p\gamma_{mn}\theta)+24\alpha'k^rk_{m}(\psi_p\gamma_{rn}\theta)+\f{3}{8}(\theta\gamma_{mn}^{\;\;\;\;\;q}\theta)\ g_{pq}-\f{3i}{4}(\theta\gamma^{tu}_{\;\;\;\;m}\theta)k_{t}b_{unp}\non\\
&&\hspace*{.4in}+\f{3}{4}\alpha'k^rk_{m}(\theta\gamma_{rn}^{\;\;\;\;\;q}\theta)\ g_{pq}
-\f{i}{24}(\theta\gamma_{tuvwmnp}\theta)k^{t}b^{uvw}-\frac{1}{6} i k_{{s}} \left(\psi _{{v}} \gamma _{{t} {u}} \theta \right) \left(\theta  \gamma _{stuv m n p} \theta \right)\non\\
&&\hspace*{.4in}-4 i \alpha  k_{{s}} k_{{t}} k_m \left(\theta  \gamma _{{t} {u} n} \theta \right) \left(\psi _p \gamma _{{s} {u}} \theta \right)+i k_{{s}} \left(\theta  \gamma _{{t} m n} \theta \right) \left(\psi _p \gamma _{{s} {t}} \theta \right)+i k_{{s}} \left(\theta  \gamma _{{t} m n} \theta \right) \left(\psi _{{t}} \gamma _{{s} p} \theta \right)\non\\
&&\hspace*{.4in}+2 i k_{{s}} \left(\theta  \gamma _{{s} {t} m} \theta \right) \left(\psi _n \gamma _{{t} p} \theta \right)-i k_{{s}} \left(\theta  \gamma _{{s} {t} m} \theta \right) \left(\psi _{{t}} \gamma _{n p} \theta\right)+\frac{1}{64 \alpha'} {(\theta \gamma_{s m n} \theta)}
{(\theta \gamma_{t u p}  \theta)}b_{s t u}\non\\
&&\hspace*{.4in}
-\frac{1}{288 \alpha'} {(\theta \gamma_{s t u} \theta)}{(\theta \gamma_{m n p}  \theta)}b_{s t u} +\frac{1}{64 \alpha'}
{(\theta \gamma_{s t u} \theta)}{(\theta \gamma_{u n p} \theta)} b_{s t m} \non\\
&&\hspace*{.4in}+\frac{1}{32} {(\theta \gamma_{s u x} \theta)}{(\theta \gamma_{t x p}  \theta)} b_{s m n } k_{t} k_{u}
-\frac{1}{16} {(\theta \gamma_{s u n} \theta)}{(\theta \gamma_{t x p}   \theta)} b_{s t m}k_{u} k_{x}\non\\
&&\hspace*{.4in}+\frac{1}{64} {(\theta \gamma_{s t x} \theta)}{(\theta \gamma_{u n p}   \theta)}b_{s t m }k_{u} k_{x}
+\frac{1}{192} {(\theta \gamma_{x z m} \theta)}{(\theta \gamma_{s t u y z n p}  \theta)}  b_{s t u}k_{x} k_{y}\non\\
&&\hspace*{.4in}+\frac{1}{192} {(\theta \gamma_{u y z} \theta)}{(\theta \gamma_{s t x z m n p}   \theta)}b_{s t u} k_{x} k_{y}
+\frac{1}{3456} {(\theta \gamma_{s t u w x y z} \theta)}{(\theta \gamma_{v x y z m n p}   \theta)} b_{s t u}k_{v} k_{w}\non\\
&&\hspace*{.4in}+\frac{1}{32} {(\theta \gamma_{s v n} \theta)}{(\theta \gamma_{t u p}  \theta)}b_{s t u}  k_{v} k_{m}
+\frac{1}{64} {(\theta \gamma_{t u v} \theta)}{(\theta \gamma_{s n p}   \theta)} b_{s t u}k_{v} k_{m}\non\\
&&\hspace*{.4in}-\frac{1}{96} {(\theta \gamma_{s t u} \theta)}{(\theta \gamma_{v n p}   \theta)} b_{s t u}k_{v} k_{m}
-\frac{1}{32} {(\theta \gamma_{s t v} \theta)}{(\theta \gamma_{u v p}   \theta)}b_{s t m} k_{u} k_{n}\non\\
&&\hspace*{.4in}+\frac{1}{384} i  \left(\theta  \gamma _{{t} {v} {w}} \theta \right) \left(\theta  \gamma _{{s} {u} {v} {w} m n p} \theta \right)k_{{u}} g_{{s} {t}}+\frac{1}{32} i  \left(\theta  \gamma _{{s} {u} n} \theta \right) \left(\theta  \gamma _{{t} {u} p} \theta \right)k_{{t}} g_{{s} m}\non\\
&&\hspace*{.4in}+\frac{1}{64} i  \left(\theta  \gamma _{{s} {t} {u}} \theta \right) \left(\theta  \gamma _{{u} n p} \theta \right)k_{{t}} g_{{s} m}+\frac{1}{64} i \left(\theta  \gamma _{{s} m n} \theta \right) \left(\theta  \gamma _{{t} {u} p} \theta \right) k_{{u}} g_{{s} {t}}\non\\
&&\hspace*{.4in}+\frac{1}{64} i  \left(\theta  \gamma _{{s} {u} m} \theta \right) \left(\theta  \gamma _{{t} n p} \theta \right)k_{{u}} g_{{s} {t}}-\frac{ i \alpha'}{16}   \left(\theta  \gamma _{{s} {u} {v}} \theta \right) \left(\theta  \gamma _{{t} {v} p} \theta \right)k_{{t}} k_{{u}} k_n g_{{s} m}\; +\ O(\theta^5)\Biggl]
\ee
where,
\be
b_{mnp} = e_{mnp} e^{ik\cdot X}\quad,\qquad e_{mn} = e_{mn} e^{ik\cdot X} \quad,\qquad \psi_{m\alpha} = e_{m\alpha} e^{ik\cdot X}
\ee

Using these theta expansion results in \eqref{vertex_ansatz} we get the theta expansion of the unintegrated vertex operator.

\section{ Three point functions using pure spinor formalism}
\label{sec:3pt_PS}
In this section, we calculate the 3-point functions involving two massless and one massive state using the pure spinor formalism. In subsection \ref{pure_cal1}, we simplify the pure spinor correlators and set up the problem at the superfield level and in subsection \ref{pure_cal2}, we evaluate the resulting correlators.

\subsectiono{Simplifying 3-point correlator } \label{pure_cal1} 
Since we are only interested in the 3-point functions on the disk, the location of all the vertex operators can be fixed and hence we only need to consider the unintegrated vertex operators. Thus, the correlator we need to evaluate is given by 
\be
\mathcal{A}_3=\bigl\langle\ V^{(1)}_A(x_1)V^{(2)}_A(x_2)V^{(3)}_{b,g,\psi}(x_3)\ \bigl \rangle \label{4.1cv}
\ee
where, $V_{b,g,\psi}$ denotes the massive vertex operator \eqref{vertex_ansatz} and $V_A$ denotes the unintegrated vertex operator of the massless states given by\footnote{The theta expansion of the massless superfields in our conventions is given in appendix \ref{appen:massless_theta}.} 
\be
V_A=\lambda^\alpha A_\alpha
\ee
The unintegrated vertex operators in \eqref{4.1cv} are fixed at some arbitrary locations $x_i$. The $SL(2,\mathbb{R})$ invariance on the disc guarantees that the 3-point function is independent of the choices of $x_i$. Using the expressions of the unintegrated vertex operators, the desired 3-point function is given by
\be
\mathcal{A}_3 &=& \ \Bigl\langle \lambda^\alpha A_\alpha^{(1)}(x_1)\lambda^\beta A_\beta^{(2)}(x_2)\Bigl(\partial \theta^\rho \lambda^\sigma B_{\sigma\rho} +d_\rho\lambda^\sigma C^\rho_{\;\sigma}+\Pi^m\lambda^\sigma H_{m\sigma}+N^{mn}\lambda^\sigma F_{\sigma mn}\Bigl)  (x_3)                                \Bigl \rangle \non\\
\label{asdf}
\ee
We have suppressed the superscript $3$ from the massive superfields since there is only one massive state and hence there is no chance of any confusion. We now manipulate each term of \eqref{asdf} one by one. 
\subsubsection*{First Term}
The first term is given by
\be
T_1&=& \ \langle \lambda^\alpha A_\alpha^{(1)}(x_1)\lambda^\beta A_\beta^{(2)}(x_2)\partial \theta^\rho \lambda^\sigma B_{\sigma\rho}  (x_3)                                 \rangle \non\\
&=&(\gamma^{mnp})_{\sigma\rho}\Bigl\langle \lambda^\alpha \tilde A_\alpha^{(1)}\lambda^\beta \tilde A_\beta^{(2)}\partial \theta^\rho \lambda^\sigma \tilde B_{mnp}                               \Bigl \rangle \left\langle e^{ik_1\cdot X(x_1)} e^{ik_2\cdot X(x_2)} e^{ik_3\cdot X(x_3)}                           \right\rangle \non\\
&=&(\gamma^{mnp})_{\sigma\rho}\Bigl\langle \lambda^\alpha \tilde A_\alpha^{(1)}\lambda^\beta \tilde A_\beta^{(2)}\partial \theta^\rho \lambda^\sigma \tilde B_{mnp}                              \Bigl \rangle \f{x_{13}x_{23}}{x_{12}}.
\ee
where, the superfields with tilde denote the same superfields with $e^{ik\cdot X}$ factor stripped off. Thus, e.g., in the theta expansion of $\tilde B_{mnp}$, we just write the polarization tensor $e_{m\alpha}, e_{mnp}$ and $e_{mn}$ instead of $\psi_{m\alpha}, b_{mnp}$ and $g_{mn}$ respectively. We have also used the momentum conservation to write
\be
 \left\langle e^{ik_1\cdot X(x_1)} e^{ik_2\cdot X(x_2)} e^{ik_3\cdot X(x_3)}  \right\rangle=|x_{12}|^{2\alpha'k_1\cdot k_2}|x_{23}|^{2\alpha'k_2\cdot k_3}|x_{13}|^{2\alpha'k_1\cdot k_3} =\f{x_{13}x_{23}}{x_{12}}.
\ee
The $T_1$ does not contribute to any of the amplitude since it does not provide the 5 zero modes of $\theta^\alpha$ which is required for the non vanishing of pure spinor correlators.

\subsubsection*{Second Term}
The 2nd term is given by
\be
T_2'= \ \langle \lambda^\alpha A_\alpha^{(1)}(x_1)\lambda^\beta A_\beta^{(2)}(x_2)d_\rho\lambda^\sigma C^\rho_{\;\sigma}  (x_3)                                 \rangle \non
\ee
To simplify this term, we use the OPE of $d_\alpha$ with superfields to obtain
\be
T_2'&=& \ \langle \lambda^\alpha A_\alpha^{(1)}(x_1)\lambda^\beta A_\beta^{(2)}(x_2)d_\rho\lambda^\sigma C^\rho_{\;\sigma}  (x_3)                                 \rangle \non\\
&=&\oint_{x_3} \f{dw}{w-x_3} \  \langle \lambda^\alpha(x_1) A_\alpha^{(1)}(x_1)\lambda^\beta(x_2) A_\beta^{(2)}(x_2)d_\rho(w)\lambda^\sigma(x_3) C^\rho_{\;\sigma}  (x_3) \rangle \non\\ 
&=&-\oint_{x_1} \f{dw}{w-x_3} \  \biggl\langle \lambda^\alpha(x_1) \biggl[ \f{\alpha'}{2}\f{D_\rho A_\alpha^{(1)}(x_1)}{w-x_1}\biggl]\lambda^\beta(x_2) A_\beta^{(2)}(x_2)\lambda^\sigma(x_3) C^\rho_{\;\sigma}  (x_3) \biggl\rangle \non\\ 
&&+\oint_{x_2} \f{dw}{w-x_3} \  \biggl\langle \lambda^\alpha(x_1)A_\alpha^{(1)}(x_1) \lambda^\beta(x_2) \biggl[ \f{\alpha'}{2}\f{D_\rho A_\beta^{(2)}(x_2)}{w-x_2}\biggl]\lambda^\sigma(z_3) C^\rho_{\;\sigma}  (x_3) \biggl\rangle \label{2.0.3}
\ee
In going to the last line, we have unwrapped the contour to enclose the points $x_1$ and $x_2$. This gives a sign. A further sign comes while moving $d_\alpha$ across $A_\alpha$. The signs in front of the individual terms in the last line are net effect of these sign factors. We now use equations \eqref{D_Bmnp_general},\eqref{2.14} and \eqref{C.11as}, the identity $QV=\f{\alpha'}{2}\lambda^\alpha D_\alpha V$ for an arbitrary superfield $V$ and unwrap the contour of $Q$. After using the on-shell condition and the pure spinor fierz identity
\be
\lambda^\alpha \lambda^\beta = \f{1}{5! \times 32} \gamma_{mnpqr}^{\alpha\beta} (\lambda\gamma^{mnpqr}\lambda)\label{2.1.8}
\ee
we obtain (after some simplification using the gamma matrix identities)
\be
T_2'=\tilde T_2 +T_2
\ee
where, 
\be
\tilde T_2&\equiv&-\f{3i\alpha'}{64}\f{x_{23}}{x_{12}}\Bigl[(\gamma_{mstuv})^{\alpha\sigma}(k_3)_m  +4(\gamma_{tuv})^{\alpha\sigma}(k_3)_s \Bigl] \Bigl\langle  \tilde A_\alpha^{(1)}\lambda^\beta \tilde A_\beta^{(2)}(\lambda\gamma^{stuvw} \lambda)\tilde \Psi_{w\sigma }                                 \Bigl\rangle\non\\[.3cm]
&&-\f{3i\alpha'}{64}\f{x_{13}}{x_{12}}\Bigl[(\gamma_{mstuv})^{\alpha\sigma}(k_3)_m  +4(\gamma_{tuv})^{\alpha\sigma}(k_3)_s \Bigl] \Bigl\langle  \lambda^\beta \tilde A_\beta^{(1)}\tilde A_\alpha^{(2)}(\lambda\gamma^{stuvw} \lambda)\tilde \Psi_{w\sigma }                                 \Bigl\rangle\label{T_2}
\ee
and 
\be
T_2&\equiv&-\f{i\alpha'}{2}\f{x_{23}}{x_{12}}(\gamma_m\gamma^{stuv})_{\alpha\sigma}(k_3)_s\  \Bigl\langle  \lambda^\alpha \tilde A^{(1)}_m\lambda^\beta \tilde A_\beta^{(2)}\lambda^\sigma \tilde B_{tuv}  \Bigl\rangle \non\\[.3cm]
&&+\f{i\alpha'}{2}\f{x_{13}}{x_{12}}(\gamma_m\gamma^{stuv})_{\alpha\sigma}(k_3)_s\  \Bigl\langle  \lambda^\beta \tilde A_\beta^{(1)}\lambda^\alpha \tilde A^{(2)}_m\lambda^\sigma \tilde B_{tuv} \Bigl\rangle \label{T_21}
\ee
\subsubsection*{Third Term}
The 3rd term is given by
\be
T_3&=& \ \langle \lambda^\alpha A_\alpha^{(1)}(x_1)\lambda^\beta A_\beta^{(2)}(x_2)\Pi^m\lambda^\sigma H_{m\sigma}  (x_3)                                 \rangle \non\\
&=&\oint_{x_3}\f{dw}{w-x_3} \langle \lambda^\alpha A_\alpha^{(1)}(x_1)\lambda^\beta A_\beta^{(2)}(x_2)\Pi^m(w)\lambda^\sigma(x_3) H_{m\sigma}  (x_3)                                 \rangle \non\\
&=&-\oint_{x_1}\f{dw}{w-x_3} \biggl\langle \lambda^\alpha(x_1) \Bigl[-i\alpha' (k_1)^m \f{A_\alpha^{(1)}(x_1)}{w-x_1}\Bigl]\lambda^\beta(x_2) A_\beta^{(2)}(x_2)\lambda^\sigma(x_3) H_{m\sigma}  (x_3)                                 \biggl\rangle \non\\
&&-\oint_{x_2}\f{dw}{w-x_3} \biggl\langle \lambda^\alpha(x_1) A_\alpha^{(1)}(x_1) \lambda^\beta(z_2) \Bigl[-i\alpha' (k_2)^m \f{A_\beta^{(2)}(x_2)}{w-x_2}\Bigl]\lambda^\sigma(x_3) H_{m\sigma}  (x_3)                                 \biggl\rangle \non\\
&=& \f{72i\alpha' }{5!\times 32}(\gamma_{stuvw})^{\alpha\sigma}(k_1)^m \Bigl\langle \tilde A_\alpha^{(1)}\lambda^\beta \tilde A_\beta^{(2)}(\lambda\gamma^{stuvw}\lambda) \tilde\Psi_{m\sigma}                              \Bigl\rangle
\label{2.1.16}
\ee
Again, in going to the 3rd equality, we have unwrapped the contour and used the OPE between $\Pi^m$ and superfields. In going to the last line, we have performed the contour integration and used the momentum conservation and the identity \eqref{2.1.8}.
\subsubsection*{Fourth Term}
Finally, the 4th term is given by
\be
T_4&=& \ \langle \lambda^\alpha A_\alpha^{(1)}(x_1)\lambda^\beta A_\beta^{(2)}(x_2)N^{mn}\lambda^\sigma F_{\sigma mn} (x_3)                                 \rangle \non\\
&=&\oint_{x_3}\f{dw}{w-z_3} \langle \lambda^\alpha A_\alpha^{(1)}(x_1)\lambda^\beta A_\beta^{(2)}(x_2)N^{mn}(w)\lambda^\sigma(x_3) F_{\sigma mn}  (x_3)                                 \rangle \non\\
&=&-\oint_{x_1}\f{dw}{w-x_3} \biggl\langle \Bigl[\f{\alpha'}{4}\f{(\gamma^{mn})^\alpha_{\ \rho}\lambda^\rho(x_1)}{w-x_1} \Bigl]A^{(1)}_\alpha(x_1) \lambda^\beta(x_2) A_\beta^{(2)}(x_2)\lambda^\sigma(x_3) F_{\sigma mn}  (x_3)                                 \biggl\rangle \non\\
&&-\oint_{x_2}\f{dw}{w-x_3} \biggl\langle \lambda^\alpha(x_1) A_\alpha^{(1)}(x_1) \Bigl[\f{\alpha'}{4}\f{(\gamma^{mn})^\beta_{\ \rho}\lambda^\rho(x_2)}{w-x_2} \Bigl]A^{(2)}_\beta(x_2)\lambda^\sigma(x_3) F_{\sigma mn}  (x_3)                                 \biggl\rangle \non\\
&=&\f{3i\alpha'}{64}\f{x_{23}}{x_{12}}\Bigl[(\gamma_{mstuv})^{\alpha\sigma}(k_3)_m  +4(\gamma_{tuv})^{\alpha\sigma}(k_3)_s \Bigl] \Bigl\langle  \tilde A_\alpha^{(1)}\lambda^\beta \tilde A_\beta^{(2)}(\lambda\gamma^{stuvw} \lambda)\tilde \Psi_{w\sigma }                                 \Bigl\rangle\non\\[.3cm]
&&+\f{3i\alpha'}{64}\f{x_{13}}{x_{12}}\Bigl[(\gamma_{mstuv})^{\alpha\sigma}(k_3)_m  +4(\gamma_{tuv})^{\alpha\sigma}(k_3)_s \Bigl] \Bigl\langle  \lambda^\beta \tilde A_\beta^{(1)}\tilde A_\alpha^{(2)}(\lambda\gamma^{stuvw} \lambda)\tilde \Psi_{w\sigma }                                 \Bigl\rangle\label{2.1.20}
\ee
In going to the 3rd equality, we have unwrapped the contour and used the OPE of $N^{mn}$ with $\lambda^\alpha$ whereas in going to the 4th equality, we have used the expression of $F_{\alpha mn}$ in terms of $\Psi_{m\alpha}$ as given in \eqref{2.14}. We note that $T_4$ is exactly minus of the $\tilde T_2$ as given in \eqref{T_2}.

\vspace*{.07in}Combining all the terms, we find that the total 3-point function is given by
\be
\mathcal{A}_3\ =\ \langle V_1V_2V_3   \rangle       \ =\  T_2+T_3\label{4.13}
\ee
where $T_2$ and $T_3$ are given in equations \eqref{T_21} and \eqref{2.1.16} respectively. {Below, we shall give the results for the correlators $\langle V_1V_2V_3   \rangle$ for different choices of the 2 massless and one massive external states.}

\subsection{Evaluation of correlators}
\label{pure_cal2}
The two terms $T_2$ and $T_3$ given in \eqref{4.13} are at the superfield level. To compute some specific 3-point function, we need to keep only the fields of interest to be non zero in the superfields. After specializing to some specific amplitude, $\mathcal{A}_3$ can be evaluated using the theta expansion results given in section \ref{thetaex} and appendix \ref{appen:massless_theta} and the pure spinor correlators listed in appendix \ref{PureSuperspace}. We use the symbolic computer programme Cadabra to do this calculation \cite{0608005,0701238}. We now give the results for different 3-point correlators. 

\subsubsection*{2 gluon and 1 $b_{mnp}$ field}
For this case, we have
\be 
T_2 &=&\f{29 i}{840}e^{m n p} e^{1}_{m} e^{2}_{n} (k_{1})_{p}\quad,\qquad 
T_3 = \frac{13 i }{840}e^{m n p} e^{1}_{m} e^{2}_{n} (k^{1})_{p}\non
\ee
This gives 
\be 
\langle aab \rangle&=& T_2+T_3\ =\ \frac{ i }{20}e^{m n p} e^{1}_{m} e^{2}_{n} (k_{1})_{p}
\ee

\subsubsection*{2 gluon and 1 $g_{mn}$ field}

For the $\langle aag\rangle$ amplitude, we have
\be 
T_2&=&-\frac{1}{80} (e^1\cdot g \cdot e^2)+\frac{\alpha'}{160} (e^2\cdot k^1)(e^1 \cdot g \cdot k^1) -\frac{\alpha'}{160} (e^1\cdot k^2)(e^2 \cdot g \cdot k^1)\non
\ee
\be
T_3&=&-\frac{3\alpha'}{160}  (e^1\cdot k^2) (e^2 \cdot g \cdot k^1) -\frac{\alpha' }{40} (e^1\cdot e^2) (k^1 \cdot g \cdot k^1)
+\frac{3 \alpha' }{160}(e^2\cdot k^1) (e^1 \cdot g \cdot k^1)\non
\ee
This gives,
\be
\langle aag \rangle&=&T_2+T_3\non\\[.2cm]
&=&-\frac{1}{80}\left[ 2\alpha'  (e^1\cdot k^2) (e^2 \cdot g \cdot k^1) +2\alpha' (e^2\cdot k^1) (e^1 \cdot g \cdot k^2)-2\alpha' (e^1\cdot e^2) (k^1 \cdot g \cdot k^2)+(e^1\cdot g\cdot e^2)\right]\non\\
\ee

\subsubsection*{2 gluino and 1 $b_{mnp}$ field}

For the $\langle\chi\chi b\rangle$ amplitude, we have
\be
T_2= \f{23}{11340}(\xi^1\gamma^{mnp}\xi^2)e_{mnp}\quad,\qquad T_3= \f{1}{18144}(\xi^1\gamma^{mnp}\xi^2)e_{mnp}\non
\ee
This gives
\be
\langle\chi\chi b\rangle\ =\ T_2+T_3&=& \f{1}{480}(\xi^1\gamma^{mnp}\xi^2)e_{mnp}  
\ee
\subsubsection*{2 gluino and 1 $g_{mn}$ field}

For the $\langle\chi\chi g\rangle$ amplitude, we have
\be
T_2= 0\quad,\qquad T_3= \f{i\alpha'}{80}(\xi^1\gamma^{m}\xi^2)g_{mn}k^n_1\non
\ee
This gives,
\be
\langle\chi\chi g\rangle\ =\ T_2+T_3&=& \f{i\alpha'}{80}(\xi^1\gamma^{m}\xi^2)e_{mn}k_1^n 
\ee

\subsubsection*{1 gluon, 1 gluino and 1 $\psi_{m\alpha}$ field}

{In this case, since all the external states are different, the two orderings $\langle a\chi\psi\rangle$ and $\langle \chi a\psi\rangle$ are different. We give the result for both cases. 
For the $\langle a\chi\psi\rangle$ correlator, we have
\be
T_2= \f{4}{21}e_1^m(\xi^2\psi_m)-\f{31}{210}\alpha' (\xi^2\psi_n)(e^1\cdot k^2)k_2^n-\f{9}{140}\alpha' (\xi^2\gamma_{mn}\psi_p)e_1^m k_1^nk_2^p\non
\ee
\be
T_3= \f{1}{105}e_1^m(\xi^2\psi_m)-\f{53}{210}\alpha' (\xi^2\psi_n)(e^1\cdot k^2)k_2^n-\f{19}{140}\alpha' (\xi^2\gamma_{mn}\psi_p)e_1^m k_1^nk_2^p\non
\ee
This gives
\be
\langle a\chi\psi\rangle&=&T_2+T_3\non\\[.2cm]
&=& \f{1}{5}\Bigl[e_1^m(\xi^2\psi_m)-2\alpha' (\xi^2\psi_n)(e^1\cdot k^2)k_2^n-\alpha'   (\xi^2\gamma_{mn}\psi_p)e_1^m k_1^nk_2^p\Bigl]\label{achipsi}
\ee
On the other hand, for the $\langle \chi a\psi\rangle$ correlator, we have
\be
T_2= \f{4}{21}e_2^m(\xi^1\psi_m)-\f{31}{210}\alpha' (\xi^1\psi_n)(e^2\cdot k^1)k_1^n-\f{9}{140}\alpha' (\xi^1\gamma_{mn}\psi_p)e_2^m k_2^nk_1^p\non
\ee
\be
T_3= \f{1}{105}e_2^m(\xi^1\psi_m)-\f{53}{210}\alpha' (\xi^1\psi_n)(e^2\cdot k^1)k_1^n-\f{19}{140}\alpha' (\xi^1\gamma_{mn}\psi_p)e_2^m k_2^nk_1^p\non
\ee
This gives
\be
\langle \chi a\psi\rangle&=&T_2+T_3\non\\[.2cm]
&=& \f{1}{5}\Bigl[e_2^m(\xi^1\psi_m)-2\alpha' (\xi^1\psi_n)(e^2\cdot k^1)k_1^n-\alpha'   (\xi^1\gamma_{mn}\psi_p)e_2^m k_2^nk_1^p\Bigl]\label{achipsi}
\ee
}

\subsubsection*{3 massless fields}
For comparing the pure spinor results with the corresponding RNS results, we also need the massless amplitudes. 
For $\langle aaa\rangle$ correlator, the pure spinor calculation gives the following result in our conventions
\be
\langle aaa \rangle&
=&\frac{i}{180}\left[  (e^1\cdot e^2) (e^3 \cdot k^1) +(e^1\cdot e^3) (e^2 \cdot k^3)+(e^2\cdot e^3) (e^1 \cdot k^2)\right]
\ee
On the other hand, for the $\langle a\chi\chi\rangle$ correlator, we get
\be
\langle a\chi \chi \rangle
&=&\frac{1}{360} (\chi^2\gamma^m \chi^3) e^{1}_m \;.
\ee

From the explicit expression of all massless-massless-massive amplitudes obtained in this section, we observe that all such 3-point functions are symmetric under a change of cyclic order. Therefore the inclusion of Chan-Paton factors (denoted by $t^a$, $t^b$ $t^c$) will make the full amplitude proportional to $ Tr(t^a,\lbrace t^b,t^c \rbrace)$. Compare this with the 3-point functions involving all massless states, which are antisymmetric under a change of cyclic order and therefore after taking into account the Chan-Paton factors, the final answer becomes proportional to $ Tr(t^a,[t^b,t^c])$. \color{black}
Another point to note is that all the amplitudes considered in this section are invariant under the gauge transformations $e^{i}\rightarrow e^{i}+k^i$. We now turn to comparing the RNS and PS results.
\section{Comparing pure spinor and the RNS results}
\label{sec:compare}
By comparing the pure spinor results given above with the corresponding RNS results given in appendix \eqref{RNS_results2}, we see that the tensor structures of the 3-point functions match perfectly. Moreover the relative coefficients of the various terms in the correlators $\langle aag\rangle$ and $\langle a\chi\psi\rangle$ which have more than one terms, also match exactly. This is a non trivial test. We shall now show that the overall numerical factors (i.e. normalizations) of the different 3-point functions in pure spinor and RNS are also in perfect agreement with each other.

\vspace*{.07in}We have denoted the overall normalization of the vertex operators in the RNS calculations relative to those in PS calculations by $g_{a}, g_\chi$ etc. (see appendix \ref{RNS_results2}). For example, if $\mathcal{N}_{RNS}$ and $\mathcal{N}_{PS}$ denote the normalizations of the gluon vertex operator in the RNS and PS respectively, then for the $\langle V_aV_aV_a\rangle$ correlator, we have (denoting $V\equiv \mathcal{N} \tilde V$)
\be
(\mathcal{N}_{PS})^3\langle \tilde V_a\tilde V_a\tilde V_a \rangle_{PS} = (\mathcal{N}_{RNS})^3\langle \tilde V_a\tilde V_a\tilde V_a \rangle_{RNS}
\ee
From this, it is clear that only the relative normalization between RNS and PS vertex operators have any physical significance. To exploit this fact, we define
\be
\mathcal{N}_{RNS} = g_a\ \mathcal{N}_{PS}
\ee
In our calculations, we have set the overall normalization of the PS vertex operators to be 1 and kept the relative normalization factor $g_a, g_\chi$ etc. to be in the RNS vertex operators.
%\be
%a_{_{RNS}} = \mathcal{N}_a \, a_{_{PS}} \quad,\quad \chi_{_{RNS}} = \mathcal{N}_{\chi} \, \chi_{_{PS}} \quad,\quad \\
%b_{_{RNS}} = \mathcal{N}_b \, b_{_{PS}} \quad,\quad g_{_{RNS}} = \mathcal{N}_g \, g_{_{PS}} \quad,\quad \psi_{_{RNS}} = \mathcal{N}_{\psi} \, \psi_{_{PS}}.
%\ee

\vspace*{.07in}With the above convention, the overall RNS and the pure spinor numerical factors for each correlator is given in the table \ref{table1}. By comparing the RNS and pure spinor numerical factors  for $\langle aaa\rangle$ and $\langle a\chi\chi\rangle$, we find 
\be
(g_a)^3=\f{-i}{180\sqrt{2\alpha'}}\quad,\qquad (g_{\chi})^2=\f{\sqrt{2}}{360\ g_a}.
\ee
\begin{table}[t]
%\begin{center}
\centering
\begin{tabular}{ | M{2cm} | M{2.6cm}| M{1.5cm}|} 
%\centering
\hline
  \small \textbf{Correlator}
& 
 \centering\small\textbf{RNS }
&
\small\textbf{PS}
\\ [.16cm]
\hline
\centering{{  \scalebox{.88}{$\langle aaa \rangle$} }}
%$t$
&
\centering{{\scalebox{.88}{$-g_a^3\sqrt{2\alpha'}$ }}}
&
\scalebox{.88}{$\f{i}{180}$}
%&
%\scalebox{.88}{$2i$}
\\ [.16cm]
\hline
\centering\scalebox{.88}{{  $\langle a\chi\chi \rangle$} }
%$t$
&
\centering{\scalebox{.88}{$\f{1}{\sqrt{2}}g_ag^2_\chi$ }}
&
\scalebox{.88}{$\f{1}{360}$}
%&
%\scalebox{.88}{$1$}
\\[.16cm]
\hline
\centering\scalebox{.88}{{  $\langle aab \rangle$} }
%$t$
&
\centering{\scalebox{.88}{$6g_a^2g_b\sqrt{2\alpha'}$ }}
&
\scalebox{.88}{$\f{i}{20}$}
%&
%\scalebox{.88}{$18i$}
 \\ [.16cm]
%\hline
\hline
\centering\scalebox{.88}{{  $\langle \chi\chi b\rangle$} }
%$t$
&
\centering{\scalebox{.88}{$-\f{1}{2\sqrt{2}}g^2_\chi g_b$ }}
&
\scalebox{.88}{$\f{1}{480}$}
%&
%\scalebox{.88}{$\f{3}{4}$}
\\ [.16cm]
\hline
\centering\scalebox{.88}{{  $\langle aag \rangle$} }
%$t$
&
\centering{\scalebox{.88}{$-g_a^2g_g$ }}
&
\scalebox{.88}{$-\f{1}{80}$}
%&
%\scalebox{.88}{$-\f{9}{2}$}
\\ [.16cm]
\hline
\centering\scalebox{.88}{{  $\langle \chi\chi g \rangle$} }
%$t$
&
\centering{\scalebox{.88}{$\sqrt{\alpha'}g_{\chi}^2g_g$ }}
&
\scalebox{.88}{$\f{i\alpha'}{80}$}
%&
%\scalebox{.88}{$\f{9i\alpha'}{2}$}
\\ [.16cm]
\hline
\centering\scalebox{.88}{{  $\langle a\chi\psi \rangle$} }
%$t$
&
\centering{\scalebox{.88}{$-\f{16\alpha'}{\sqrt{2}}g_ag_{\chi} g_{\psi}$ }}
&
\scalebox{.88}{$\f{1}{5}$}
%&
%\scalebox{.88}{{$72$}}
\\ [.16cm]
\hline
\centering\scalebox{.88}{{  $\langle \chi a\psi \rangle$} }
%$t$
&
\centering{\scalebox{.88}{$-\f{16\alpha'}{\sqrt{2}}g_ag_{\chi} g_{\psi}$ }}
&
\scalebox{.88}{$\f{1}{5}$}
%&
%\scalebox{.88}{{$72$}}
\\ [.16cm]
\hline
\end{tabular}
\caption{\small Numerical factors of RNS and pure spinor correlators.}\label{table1}
%\end{center}
\end{table}
In terms of $g_a$ and $g_{\chi}$, the numerical factors for $\langle aab\rangle$, $\langle aag\rangle$ and $\langle a\chi\psi   \rangle$ give
\be
g_b =  \f{i}{120\sqrt{2\alpha'}\ g_a^2}\quad,\qquad g_g =  \f{1}{80\ g_a^2}\quad,\qquad g_{\psi} = -\f{\sqrt{2}}{80\alpha' g_ag_{\chi}}
\ee
The above values of $g_b,g_g$ and $g_\psi$ agree perfectly with the value obtained using $\langle \chi\chi b\rangle$, $\langle \chi\chi g\rangle$ and $\langle \chi a \psi \rangle$ correlators. This is a non trivial consistency check.

\section{Discussions} \label{outlook}

In this paper we have explicitly shown the equivalence of all massless-massless-massive 3-point functions for each given ordering of vertex operators on the boundary \color{black} for open superstrings computed in PS and RNS formalisms. For the  massive states, the normalization factors, which are fixed, using say $\langle aab\rangle$, $\langle aag\rangle$ and $\langle a\chi \psi \rangle$ correlators, match perfectly with the $\langle \chi\chi b\rangle$, $\langle \chi\chi g\rangle$ and $\langle \chi a \psi \rangle$ correlators which establishes the consistency of the relative normalization determined in this paper and establishes firmly the equivalence between RNS and PS formalism for first massive states. \color{black}

\vspace*{.07in}Once the vertex for open superstring is known, its extension to closed strings can be obtained in a straightforward manner by taking appropriate tensor products of left and right moving sectors \cite{heterotic}. 

\vspace*{.07in}To compare loop amplitudes one needs to compute the amplitude in PS formalism using integrated vertex constructed for massive states in \cite{integrated}. With the relative normalization now fixed, they should match exactly with the corresponding RNS result. Also as argued in \cite{integrated}, the strategy behind constructing massive vertex (integrated or unintegrated) is not sensitive to mass level in question and is expected to be identical for all higher massive states. Therefore, one can reasonably expect that all 3-point functions involving higher massive states can also be directly compared with RNS results and yield a consistent set of normalizations for each mass level.

\bigskip

\bigskip

\noindent{\bf Acknowledgments:}  
We are deeply thankful to Ashoke Sen for many insightful discussions throughout the course of this work and for useful comments on the draft. We would also like to thank Anirban Basu, Rajesh Gopakumar and Satchitananda Naik for their encouragement and comments on the draft.  SPK is also thankful to the
organizers of Strings 2018 held in Okinawa Institute of Science and Technology (OIST), Japan for giving him the opportunity to give a gong show on a preliminary version of this
work. MV is also thankful to Institute of Theoretical Physics, Chinese Academy of Sciences, China and the OIST, Japan for hospitality while this work was in progress. This research was supported in part by the Infosys Fellowship for the senior students. We also thank the people and Government of India for their continuous support for theoretical physics. 
\appendix
\section{Some pure spinor results}
\label{appen:PS}
In this appendix, we note some pure spinor results which are used in this work. 

\subsection{Pure spinor conventions}
In this section, we give our conventions for the pure spinor calculations. For the tree amplitude calculations, both the minimal as well as the non minimal formalisms give the same prescription. For convenience, we shall only consider the minimal formalism. The world sheet action of the minimal formalism is given by
\be
S=\f{1}{\pi \alpha'}\int d^2z \left(\f{1}{2}\p X^m\bar\p X_m+p_\alpha\bar\p\theta^\alpha-w_\alpha\bar\p\lambda^\alpha\right)
\ee
where, $m=0,1,,\cdots,9$ and $\alpha=1,\cdots,16$. 

\vspace*{.07in}The $p_\alpha$ and $ w_\alpha$ are the conformal weight one conjugate momenta of the conformal weight zero fields $\theta^\alpha$ and $ \lambda^\alpha$ respectively. The fields with upper (lower) spinor indices transform as right (left) handed Weyl spinor. The $\lambda^\alpha$ satisfy the pure spinor constraint
\be
\lambda^\alpha\gamma^{m}_{\alpha\beta}\lambda^\beta=0\label{psconstr}
\ee
The BRST operator is defined to be
\be
Q=\oint dz\ \bigl(\lambda^\alpha d_\alpha\bigl)
\ee
Due to pure spinor constraint, the world-sheet fields can only appear in the following gauge invariant combinations\footnote{The coincident operators are normal ordered via
\be 
:A(z)B(z): \equiv \frac{1}{2 \pi i} \oint_z  \frac{dw}{w-z} A(w)B(z)  \label{normal_order}
\ee
 where, $A$ and $B$ are any two operators and the contour surrounds the point $z$. However, we shall suppress the normal ordering symbol $:\ :$ throughout this draft.}
\be
N_{mn}=\f{1}{2} w_\alpha(\gamma_{mn})^\alpha_{\;\beta}\lambda^\beta\quad,\quad J=w_\alpha \lambda^\alpha\quad,\quad T= w_\alpha \p\lambda^\alpha\non
\ee
Any other gauge invariant combination can be expressed in terms of these objects. The ghost current is given by $J= \lambda^\alpha w_\alpha $ which implies that the $\lambda^\alpha$ carries the ghost number $+1$ and the rest of the fields carry zero ghost number.  

\vspace*{.07in}Before proceeding ahead, we note down some important OPEs of the theory which arise frequently in calculations 
\be
d_\alpha(z)d_\beta(w)=-\f{\alpha'\gamma^m_{\alpha\beta}}{2(z-w)}\Pi_m(w)+\cdots
\quad,\qquad
d_\alpha(z)\Pi^m(w)=\f{\alpha'\gamma^m_{\alpha\beta}}{2(z-w)}\partial\theta^\beta(w)+\cdots\non
\ee
\be
d_\alpha(z)V(w)=\f{\alpha'}{2(z-w)}D_\alpha V(w)+\cdots \quad,\qquad
\Pi^m(z)V(w)=-\f{\alpha'}{(z-w)}\partial^mV(w)+\cdots\non
\ee
\be
 \Pi^m(z)\Pi^n(w)=-\f{\alpha'\eta^{mn}}{2(z-w)^2}+\cdots\quad,\qquad N^{mn}(z)\lambda^\alpha(w)=\f{\alpha'(\gamma^{mn})^\alpha_{\;\;\beta}}{4(z-w)}\ \lambda^\beta(w)+\cdots
\non
\ee
\be
J(z)J(w)=-\f{(\alpha')^2}{(z-w)^2}+\cdots\quad,\qquad  J(z)\lambda^\alpha(w)=\f{\alpha'}{2(z-w)}\lambda^\alpha(w)+\cdots\non
\ee
\be
N^{mn}(z)N^{pq}(w)=-\f{3(\alpha')^2}{2(z-w)^2}\eta^{m[q}\eta^{p]n}-\f{\alpha'}{(z-w)}\Bigl(\eta^{p[n}N^{m]q}-\eta^{q[n}N^{m]p}\Bigl)+\cdots
\ee
In the above OPEs, $\p_m$ is the derivative with respect to the spacetime coordinate $X^m$, $\p$ is the derivative with respect to the world-sheet coordinate, $V$ denotes an arbitrary superfield. The $d_\alpha$ and $\Pi^m$ denote the supersymmetric combinations
\be
d_\alpha&=&p_\alpha-\f{1}{2}\gamma^m_{\;\;\alpha\beta}\theta^\beta\partial X_m-\f{1}{8}\gamma^m_{\alpha\beta}\gamma_{m\sigma\delta}\theta^\beta\theta^{\sigma}\partial\theta^\delta
\non\\[.4cm]\Pi^m&=&\partial X^m+\f{1}{2}\gamma^m_{\alpha\beta}\theta^\alpha\partial\theta^\beta
\ee
The $D_\alpha$ is the supercovariant derivative given by
 \be D_\alpha\equiv\p_\alpha+\gamma^m_{\alpha\beta}\theta^\beta\p_m
\quad\implies\qquad
\lbrace D_\alpha, D_\beta\rbrace= 2 (\gamma^m)_{\alpha \beta} \partial_m\label{cov_der}
\ee
For calculating the scattering amplitudes, we need to know the number of zero modes of the world-sheet fields. Due to the constraints \eqref{psconstr} and the gauge freedom \eqref{gauge_trans}, it follows that all the worlds-heet ghost sector fields, namely $\lambda^\alpha$ and $ w_\alpha$ have 11 independent components. Now, it is a result from the theory of Riemann surfaces that the number of zero modes of the conformal weight one and conformal weight zero objects on a genus $g$ Riemann surface is $g$ and $1$ respectively. Thus, the number of zero modes of $ \lambda^\alpha$ on a genus $g$ Riemann surface is 11 whereas the number of zero modes of $w_\alpha$ on a genus $g$ Riemann surface is $11g$. On the other hand, the matter sector fields $\theta^\alpha$ and $p_\alpha$ have $16$ and $16g$ zero modes respectively on a genus $g$ Riemann surface. As in RNS formalism, the integration over the zero modes give divergences and we need some way to absorb these zero modes.

\subsection{Hodge duality of gamma matrices in d=10}

In $10$ dimensions, the $16\times16$ gamma matrices satisfy the following Hodge dualities
\be 
(\gamma^{m_1...m_{2n}})^{\alpha}_{\; \; \beta} &=& \frac{1}{(10-2n)!}(-1)^{(n+1)} \epsilon^{m_1...m_{2n} p_1...p_{10-2n}}(\gamma_{p_1...p_{10-2n}})^{\alpha}_{\; \; \beta}
\ee
\be 
(\gamma^{m_1...m_{2n}})_{\alpha}^{\; \; \beta} &=& -\frac{1}{(10-2n)!}(-1)^{(n+1)} \epsilon^{m_1...m_{2n} p_1...p_{10-2n}}(\gamma_{p_1...p_{10-2n}})_{\alpha}^{\; \; \beta}
\ee
\be 
(\gamma^{m_1...m_{2n+1}})^{\alpha \beta} &=& \frac{1}{(9-2n)!}(-1)^{n} \epsilon^{m_1...m_{2n+1} p_1...p_{9-2n}}(\gamma_{p_1...p_{9-2n}})^{\alpha \beta}
\ee
\be 
(\gamma^{m_1...m_{2n+1}})_{\alpha \beta} &=& -\frac{1}{(9-2n)!}(-1)^{n} \epsilon^{m_1...m_{2n+1} p_1...p_{9-2n}}(\gamma_{p_1...p_{9-2n}})_{\alpha \beta}
\ee
where, $\epsilon^{m_1\cdots m_{9}}$ is the 10 dimensional epsilon tensor defined as
\be
\epsilon_{0\;1\;\cdots\;9}=1\qquad\implies \qquad \epsilon^{0\;1\;\cdots\;9}=-1
\ee
Due to the above dualities, not all the antisymmetrized product of gamma matrices are independent. We shall take $\gamma^{m_1}$, $\gamma^{m_1 m_2}$, $\gamma^{m_1 m_2 m_3}$, $\gamma^{m_1 m_2 m_3 m_4 }$ and $\gamma^{m_1 m_2 m_3 m_4 m_5}$ 
along with the identity matrix $\mathbb{I}_{16 \times 16}$ as the linearly independent basis elements for vector spaces of $16 \times 16$ complex matrices.

\subsection{Pure spinor superspace identities}  \label{PureSuperspace}
While computing the scattering amplitudes in pure spinor formalism, the last step requires evaluation of integration over the zero modes of $\lambda^\alpha$ and $\theta^\alpha$. Due to the pure spinor constraints and the symmetry properties of $\theta^\alpha$ and $\lambda^\alpha$, there are only a finite number of these basic pure spinor correlators which are non zero. Below, we list those pure spinor correlators which are used in this note \cite{Berkovits_Mafra}
\be
\langle(\lambda\gamma^m\theta)(\lambda\gamma^n\theta)(\lambda\gamma^p\theta)(\theta\gamma_{stu}\theta)\rangle=\f{1}{120}\delta^{mnp}_{stu}
\ee
\be
\langle(\lambda\gamma^{pqr}\theta)(\lambda\gamma_m\theta)(\lambda\gamma_n\theta)(\theta\gamma_{stu}\theta)\rangle=\f{1}{70}\delta^{[p}_{[m}\eta_{n][s}\delta^q_{t}\delta^{r]}_{u]}
\ee
\be
\langle(\lambda\gamma^{mnpqr}\theta)(\lambda\gamma_s\theta)(\lambda\gamma_{t}\theta)(\theta\gamma_{uvw}\theta)\rangle=-\f{1}{42}\delta^{mnpqr}_{stuvw}-\f{1}{5040}\epsilon^{mnpqr}_{\;\;\;\;\;\;\;\;\;\;\;\; stuvw}
\ee

\be
\langle(\lambda\gamma_{q}\theta)(\lambda\gamma^{mnp}\theta)(\lambda\gamma^{rst}\theta)(\theta\gamma_{uvw}\theta)\rangle
&=&-\f{1}{280}\Bigl[\eta_{q[u}\eta^{z[r}\delta^{s}_{v}\eta^{t][m}\delta^n_{w]}\delta^{p]}_{z}-\eta_{q[u}\eta^{z[m}\delta^{n}_{v}\eta^{p][r}\delta^s_{w]}\delta^{t]}_{z}\Bigl]\non\\
&&+\f{1}{140}\Bigl[\delta^{[m}_{q}\delta^{n}_{[u}\eta^{p][r}\delta^s_{v}\delta^{t]}_{w]}-\delta^{[r}_{q}\delta^{s}_{[u}\eta^{t][m}\delta^n_{v}\delta^{p]}_{w]}\Bigl]\non\\
&&-\f{1}{8400}\epsilon^{qmnprstuvw}
\ee
\be
&&\hspace*{-.95in}\langle(\lambda\gamma^{mnpqr}\theta)(\lambda\gamma_{stu}\theta)(\lambda\gamma^{v}\theta)(\theta\gamma_{wxy}\theta)\rangle\non\\
&=&\f{1}{120}\epsilon^{mnpqr}_{\;\;\;\;\;\;\;\;\;\;\;\;ghijk}\left(\f{1}{35}\eta^{v[g}\delta^{h}_{[s}\delta^{i}_{t}\eta_{u][w}\delta^j_{x}\delta^{k]}_{y]}-\f{2}{35}\delta^{[g}_{[s}\delta^{h}_{t}\delta^{i}_{u]}\delta^{j}_{[w}\delta^{k]}_{x}\delta^{v}_{y]}\right)\non\\
&&+\f{1}{35}\eta^{v[m}\delta^{n}_{[s}\delta^{p}_{t}\eta_{u][w}\delta^q_{x}\delta^{r]}_{y]}-\f{2}{35}\delta^{[m}_{[s}\delta^{n}_{t}\delta^{p}_{u]}\delta^{q}_{[w}\delta^{r]}_{x}\delta^{v}_{y]}
\ee

\section{RNS results}
\label{appen:RNS_results}
In this section, we summarize the results of the 3-point functions involving two massless and one massive states computed using the RNS formalism. In subsection \ref{RNS_convention}, we state our conventions and in subsection \ref{RNS_results2}, we give the results of 3-point computations. A useful reference for this section is \cite{Hartl:2010ks} (also see \cite{Oliver1})
\subsection{Conventions for RNS calculations} 
\label{RNS_convention}

Due to the picture number anomaly on a genus $g$ Riemann surface, a non vanishing RNS correlator must have the picture number $2g-2$. This is ensured by working with vertex operators in appropriate picture number and the insertions of appropriate number of picture changing operators (PCOs). For the 3-point functions on Riemann sphere, we can avoid the insertions of PCOs by working with vertex operators of the appropriate picture number so that the total picture number adds up to $-2$ (which is the picture number anomaly on Riemann sphere). 

\vspace*{.07in}We start by writing down the vertex operators for the massless states in various picture numbers. The gluon vertex operator in the $-1$ and $0$ picture numbers is given by
\be
V_{a}^{(-1)}(x)&=&g_a\ e_m \psi^m(x)e^{-\phi(x)} e^{ik\cdot X(x)}\non\\[.4cm]
V_{a}^{(0)}(x)&=&\f{g_a}{\sqrt{2\alpha'}}e_m \Bigl(i\p X^m(x) +2\alpha'k_n\psi^n(x)\psi^m(x)\Bigl)e^{ik\cdot X(x)}
\ee 
The gluino vertex operator in the $-1/2$ picture number is given by
\be
V^{(-1/2)}_\chi(x)\ &=&\ g_{\chi}\ \xi^\alpha S_\alpha(x) e^{-\phi(x)/2} e^{ik\cdot X(x)} 
%V^{(-3/2)}_\chi(z)\ &=&\ \f{\mathcal{N}_{\chi}}{\sqrt{\alpha'}} \bar v_{\dot\beta} S^{\dot\beta}(z) e^{-3\phi(z)/2} e^{ik\cdot X(z)} \non\\[.3cm]
%V^{(1/2)}_\chi(z)\ &=&\ \f{\mathcal{N}_{\chi}}{2\sqrt{\alpha'}} \xi^\alpha\Bigl(i\p X_m(z)+\f{\alpha'}{2}k_n\psi^n(z)\psi_m(z)\Bigl) \gamma^m_{\alpha\dot\beta}S^{\dot\beta}(z) e^{\phi(z)/2} e^{ik\cdot X(z)}
\ee
The polarization vectors of the gluon and gluino satisfy the transversality conditions
\be
e_m k^m\ =\ 0\quad,\qquad\xi^\alpha \gamma^m_{\alpha\dot\beta}k_m\ =\ 0
\ee
Now, we turn to the vertex operators for the first massive states \cite{Koh:1987hm, Tanii:1987bk}. The vertex operators for the anti-symmetric 3-form field $b_{mnp}$ in the picture numbers $-1$ and $0$ are given by
\be
V_{b}^{(-1)}(x)&=& g_b\ e_{mnp}\psi^m(x)\psi^n(x)\psi^p(x)e^{-\phi(x)} e^{ik\cdot X(x)}\non\\[.3cm]
V_{b}^{(0)}(x)&=&  g_b\sqrt{2\alpha'}\ e_{mnp}\Bigl(\f{3i}{2\alpha'}\p X^m \psi^n\psi^p+k_q\psi^q\psi^m\psi^n\psi^p\Bigl)e^{ik\cdot X(x)}
\ee
The vertex operators for the symmetric traceless massive graviton field $g_{mn}$ are given by
\be
V_{g}^{(-1)}(x)&=&\f{g_g}{\sqrt{2\alpha'}}\ e_{mn}i\p X^m(x)\psi^n(x)e^{-\phi(x)} e^{ik\cdot X(x)}\non\\[.3cm]
V_{g}^{(0)}(x)&=& g_g\ e_{mn}\Bigl(\f{1}{2\alpha'}i\p X^m i\p X^n+\p\psi^m\psi^n+i\p X^mk_p\psi^p\psi^n\Bigl)e^{ik\cdot X}
\ee
Finally, the vertex operators for the massive gravitino field $\psi_{m\alpha}$ in the $-1/2$ picture number is given by
\be
V_{\psi}^{(-1/2)}(z)&=& \f{g_{\psi}}{\sqrt{\alpha'}} \Bigl(\xi^\alpha_mi\p X^m-\f{\alpha'}{4}\xi_m^\beta\psi^m\gamma^n_{\beta\dot\beta}\gamma_p^{\dot\beta\alpha}k_nk^p\Bigl)S_\alpha e^{-\phi/2}e^{ik\cdot X}(z)
\ee
The polarization vectors of the 1st massive states satisfy the conditions
\be
k^m e_{mnp} = k^m e_{mn} = k^m\xi^\alpha_m = 0
\ee
For comparison with PS result, it is useful to parametrize the massive tensor spinor $\xi^\alpha_m$ as
\be
\xi_m^\alpha = -8\alpha' \bar\rho_{m\dot\beta}k^p\bar\gamma^{\dot\beta\alpha}_p\qquad,\qquad k^m\bar\rho_{m\dot\beta}=\bar\rho_{m\dot\beta}\gamma_m^{\dot\beta\alpha}=0 
\ee

We now turn to the OPEs and correlation functions of the world-sheet matter and ghost sector fields. The important correlators of the open string $X^m$ fields are given by 
\be
\left\langle \prod_{j=1}^n e^{ik_j\cdot X(y_j)} \right\rangle\ =\ \prod_{i<j}^n |y_i-y_j|^{2\alpha'k_i\cdot k_j}\label{mar_corr1}
\ee

\be
\left\langle\prod_{\ell=1}^q i\p X^{m_\ell}(w_\ell) \prod_{j=1}^n e^{ik_j\cdot X(y_j)} \right\rangle\ &=& \sum_{i=2}^p\f{2\alpha'\eta^{m_1m_i}}{(w_1-w_i)^2}\left\langle \prod_{\ell=2\atop \ell\not=i}^p i\p X^{m_\ell}(w_\ell)\prod_{j=1}^ne^{ik_j\cdot X(y_j)} \right\rangle \non\\
&& + \sum_{\ell=1}^n\f{2\alpha'k_\ell^{m_1}}{(w_1-y_\ell)}\left\langle \prod_{\ell=2}^p i\p X^{m_\ell}(w_\ell)\prod_{j=1}^ne^{ik_j\cdot X(y_j)} \right\rangle\label{mat_corr2}
\ee
This can be evaluated recursively using \eqref{mar_corr1}. We shall encounter the situation where some $w_\ell$ may coincide with some $y_j$. E.g., we shall need the following correlator
\be
\Gamma^m&\equiv&\left\langle e^{ik_1\cdot X(z_1)}e^{ik_2\cdot X(z_2)}\p X^m(z_3)e^{ik_3\cdot X(z_3)} \right\rangle\non\\
&=&\oint \f{dw}{w-z_3}\left\langle e^{ik_1\cdot X(z_1)}e^{ik_2\cdot X(z_2)}e^{ik_3\cdot X(z_3)}\p X^m(w) \right\rangle\non\\
%&=&-i\oint \f{dw}{w-z_3}\left(\f{2\alpha'k_1^m}{w-z_1}+\f{2\alpha'k_2^m}{w-z_2}+\f{2\alpha'k_3^m}{w-z_3}\right)|z_{1}-z_2|^{2\alpha'k_1\cdot k_2}|z_2-z_3|^{2\alpha'k_2\cdot k}|z_{1}-z_3|^{2\alpha'k_1\cdot k}\non\\
&=&\left(\f{2i\alpha'k_1^m}{z_{13}}+\f{2i\alpha'k_2^m}{z_{23}}\right)|z_{12}|^{2\alpha'k_1\cdot k_2}|z_{23}|^{2\alpha'k_2\cdot k_3}|z_{13}|^{2\alpha'k_1\cdot k_3}
%&=&-\left(\f{2i\alpha'k_1^m}{z_3-z_1}+\f{2i\alpha'k_2^m}{z_3-z_2}\right)(z_{1}-z_2)^{-1}(z_2-z_3)(z_{1}-z_3)\non\\
%&=&2i\alpha'\left(\f{k_1^m z_{23}+ k^m_2 z_{13}}{z_{12} }\right)
\ee
In going to the 2nd line, we have used the definition of the normal ordering \eqref{normal_order} and in going to the 3rd line, we have used \eqref{mat_corr2} for $q=1$ and $n=3$. 

\vspace*{.07in}Similarly, we also need 
\be
&&\Bigl\langle e^{ik_1\cdot X(z_1)}   e^{ik_2\cdot X(z_2)}i\p X^p(z_3)i\p X^q(z_3)e^{ik_3\cdot X(z_3)}\Bigl\rangle\non\\[.3cm]
&=& 4(\alpha')^2\left[\f{k_1^p}{z_{13}}\biggl(\f{k_1^q}{z_{13}}+\f{k_2^q}{z_{23}}\biggl)\ +\  \f{ k_2^p}{z_{23}}\biggl(\f{ k_1^q}{z_{13}}+\f{k_2^q}{z_{23}}\biggl)\right]|z_{12}|^{2\alpha'k_1\cdot k_2}|z_{23}|^{2\alpha'k_2\cdot k_3}|z_{13}|^{2\alpha'k_1\cdot k_3}
      \non
\ee
Next, we consider the world-sheet correlators involving the $\psi^m$ fields
\be
\left\langle \prod_{i=1}^n \psi^{m_i} (y_i) \right\rangle\ &=& \sum_{j=2}^n(-1)^j\f{\eta^{m_1m_j}}{(y_1-y_j)}\left\langle \prod_{\ell=2 \atop \ell\not=j}^n \psi^{m_\ell}(y_\ell) \right\rangle %\label{mat_corr3}
\ee
This expression can also be evaluated recursively. In the final step, the two point function can be evaluated by using the OPE
\be
\psi^m(z)\psi^n(w) \ =\ \f{\eta^{mn}}{z-w}+\cdots %\label{OPEpsi}
\ee 

\vspace*{.07in}Next, we consider the ghost sector. The basic correlator involving the reparametrization ghost $c(x)$ is given by
\be
\langle c(x_1)c(x_2)c(x_3) \rangle \ =\ x_{12}x_{23} x_{13}
\ee
The basic correlator involving the bosonized ghost field $\phi(z)$ is given by
\be
\left\langle \prod_{k=1}^n e^{q_k\phi(z_k)}     \right\rangle\ =\ \prod_{k < \ell}^n \f{1}{(z_{k}-z_{\ell})^{q_kq_\ell}} \qquad;\qquad \sum_{k=1}^n q_k \ =\ -2
\ee
Finally, we consider the spin fields. The basic OPEs involving the spin fields are given by (a consistent set of these OPEs can be found in \cite{Sen:2016bwe}) 
\be
\psi^m(z)S_\alpha(w)e^{-\phi(w)/2}\ =\ \f{(\bar\gamma^m)^{\dot\alpha\beta}S_{\beta}e^{-\phi(w)/2}}{\sqrt{2}(z-w)^{1/2}} +\cdots
\ee
\be
\psi^m(z)S^{\dot\alpha}(w)e^{-\phi(w)/2}\ =\ \f{(\gamma^m)_{\alpha\dot\beta}S^{\dot\beta}e^{-\phi(w)/2}}{\sqrt{2}(z-w)^{1/2}} +\cdots
\ee
\be
S_\alpha(z)e^{-\phi(z)/2}S^{\dot\beta}(w)e^{-3\phi(w)/2}\ =\ \f{C_{\alpha}^{\;\;\dot\beta}e^{-2\phi(w)}}{(z-w)^{2}} +\cdots
\ee
\be
S_\alpha(z)e^{-\phi(z)/2} S_\beta(w)e^{-\phi(w)/2}\ =\ \f{(\gamma_mC)_{\alpha\beta}\psi^m(w)e^{-\phi(w)}}{\sqrt{2}(z-w)}+\cdots
\ee
\be
\psi_m(z)e^{-\phi(z)} S_\alpha(w)e^{-\phi(w)/2}\ =\ \f{(\bar\gamma_m)_{\alpha\dot{\beta}}S^{\dot\beta}(w)e^{-3\phi(w)/2}}{\sqrt{2}(z-w)}+\cdots
\ee

\be
j^{mn}(z)S_\alpha(w) \ =\  -\f{(\gamma^{mn})_\alpha^{\;\beta}S_\beta(w)}{2(z-w)}\quad,\qquad j^{mn}\ \equiv\ :\psi^m\psi^n:
\ee
Using the above OPEs, we can work out the following useful correlators which are needed in our calculations
\be
\langle\psi^m(x_1)e^{-\phi(x_1)}S_\alpha(x_2)e^{-\phi(x_2)/2}S_{\beta}(x_3)e^{-\phi(x_3)/2}\rangle = \f{(\gamma^m C)_{\alpha\beta}}{\sqrt{2}z_{12}z_{13}z_{23}}
\ee
\be
\xi^{\beta}_q\Bigl\langle \psi^m(x_1)e^{-\phi(x_1)}S_\alpha(x_2)e^{-\phi(x_2)/2}\psi^q(x_3)\psi^p(x_3)\gamma_p^{\dot\beta\sigma}S_\sigma(x_3)e^{-\phi(x_3)/2}  \Bigl\rangle = \f{8\ \xi^{\beta}_mC_\alpha^{\;\dot\beta}}{\sqrt{2}z_{13}^{2}z_{23}^{2}}
\ee

\subsection{3-point functions} 
\label{RNS_results2}

We are now ready to give the results of 3-point functions of two massless and one massive field (for the computation of 3-point functions of states in leading Regge trajectory in RNS, see \cite{Schlotterer:2010kk}). These are straightforward to evaluate using the vertex operators and the correlators given in the previous subsection. Hence, we just state the final results below. For our calculations, we shall take the bosonic massless and massive fields to be either in the $-1$ or $0$ picture numbers. On the other hand, the fermionic massless or massive fields will always be taken in the $-1/2$ picture. The picture numbers will be shown by a superscript on the vertex operators inside correlators. Thus, the 3-point amplitudes are given by
\be
\mathcal{A}_3=\bigl\langle\ c(x_1)V^{(1)}_{a,\chi}(x_1)c(x_2)V^{(2)}_{a,\chi}(x_2)c(x_3)V^{(3)}_{b,g,\psi}(x_3)\  \rangle\label{4.1}
\ee

\vspace*{.07in}We start by considering the 3-point functions involving all massless fields. The two possible correlators in this case are $\langle aaa\rangle$ and $\langle a\chi\chi\rangle$. Using the world-sheet correlators given above, these 3-point functions can be evaluated to be
\be
\langle aaa \rangle&\equiv& \left\langle c(x_1)V_a^{(-1)}(x_1)c(x_2)V_a^{(-1)}(x_2)c(x_3)V_a^{(0)}(x_3) \right\rangle \non\\[.2cm]
&=&-g_a^3\sqrt{2\alpha'}\Bigl[(e^1\cdot e^2)(e^3\cdot k^1 )+(e^1\cdot e^3)(e^2\cdot k^3)+(e^2\cdot e^3)(e^1\cdot k^2)\Bigl] %\label{aaa}
\ee
As an amplitude this vanishes on summing $(k_1\leftrightarrow k_2)$ term if the gauge group is abelian. 

\vspace*{.07in}Similarly, the $a\chi\chi$ correlator can be worked out to be
\be
\langle a\chi\chi \rangle&\equiv& \left\langle c(x_1)V_a^{(-1)}(x_1)c(x_2)V_\chi^{(-1/2)}(x_2)c(x_3)V_\chi^{(-1/2)}(x_3) \right\rangle\ \non\\[.2cm]
&=&\f{1}{\sqrt{2}}g_ag_{\chi}^2e^{1}_m\Bigl(\xi^2\gamma^mC\xi^3\Bigl)%\label{achichi}
\ee
Again, for the abelian gauge group, the corresponding amplitude vanishes. 

\vspace*{.07in}Next, we consider the two massless and one massive field. We start with two gluons and one $b_{mnp}$ field
\be
\langle aab \rangle&\equiv& \left\langle c(x_1)V_a^{(-1)}(x_1)c(x_2)V_a^{(-1)}(x_2)c(x_3)V_b^{(0)}(x_3) \right\rangle \non\\[.2cm]
&=&6g_a^2 g_b\sqrt{2\alpha'}e_m^{1}e_n^{2}e_{mnp}k_1^p
\ee

\vspace*{.07in}Next, we consider the 3-point function of two gluons and one massive $g_{mn}$ field which is given by
\be
\langle aag \rangle&\equiv& \left\langle c(x_1)V_a^{(-1)}(x_1)c(x_2)V_a^{(-1)}(x_2)c(x_3)V_g^{(0)}(x_3) \right\rangle\non\\[.3cm]
&=&-g_a^2g_g\Bigl[ 2\alpha' \eta^{mn}e^{1}_me^{2}_ne_{pq}k_1^p k_1^q +e^{1}_pe^{2}_qe_{pq}+2\alpha'e^{1}_me^{2}_qe_{pq}k_1^p k_2^m- 2\alpha'e^{1}_qe^{2}_me_{pq}k_1^m k_1^p\Bigl]  \non\\
\ee

Next, we include gluino and consider the $\langle\chi\chi b\rangle$ and $\langle\chi\chi g\rangle$ correlators which can be worked out to be
\be
\langle \chi\chi b \rangle&\equiv& \left\langle c(x_1)V_\chi^{(-1/2)}(x_1)c(x_2)V_\chi^{(-1/2)}(x_2)c(x_3)V_{b}^{(-1)}(x_3) \right\rangle\non\\
&=&-\f{1}{2\sqrt{2}}g^2_\chi g_b (\xi^{1}\gamma^{mnp}C\xi^{2})e_{mnp}
\ee
and,
\be
\langle \chi\chi g \rangle&\equiv& \langle c(x_1)V_\chi^{(-1/2)}(x_1)c(x_2)V_\chi^{(-1/2)}(x_2)c(x_3)V_{g}^{(-1)}(x_3) \rangle\non\\
&=&\sqrt{\alpha'}g^2_\chi g_g e_{mn}(\xi^{1}\gamma^nC\xi^{2}) k_1^m 
\ee
Finally, we consider the massive fermion. The two different 3-point functions with one massive fermion and two massless fields are $\langle a\chi\psi\rangle$ and $\langle \chi a\psi\rangle$ which are given by
\be
\langle a\chi\psi \rangle&\equiv& \left\langle c(x_1)V_a^{(-1)}(x_1)c(x_2)V_\chi^{(-1/2)}(x_2)c(x_3)V_{\psi}^{(-1/2)}(x_3) \right\rangle\non\\[.2cm]
&=&-\f{16\alpha'}{\sqrt{2}}g_ag_{\chi} g_{\psi}  \biggl[ -\alpha'e_m^{(1)}\xi_{(2)}^\alpha  k_2^qk^p_1\bar\rho_{q\dot\beta}(\bar\gamma_{mp})^{\dot\beta}_{\;\;\dot\sigma} C^{\dot\sigma}_{\;\;\alpha}-2\alpha'k_2^qk^m_2e_m^{(1)}\xi_{(2)}^\alpha\bar\rho_{q\dot\beta}C^{\dot\beta}_{\;\;\alpha}+e_m^{(1)}\xi_{(2)}^\alpha\bar\rho_{m\dot\sigma}C_{\;\alpha}^{\dot\sigma}\biggl]\non\\
\label{achipsiPS}
\ee
and,
\be
\langle \chi a\psi \rangle&\equiv& \left\langle c(x_1)V_\chi^{(-1/2)}(x_1)c(x_2)V_a^{(-1)}(x_2)c(x_3)V_{\psi}^{(-1/2)}(x_3) \right\rangle\non\\[.2cm]
&=&-\f{16\alpha'}{\sqrt{2}}g_ag_{\chi} g_{\psi}  \biggl[ -\alpha'e_m^{(2)}\xi_{(1)}^\alpha  k_1^qk^p_2\bar\rho_{q\dot\beta}(\bar\gamma_{mp})^{\dot\beta}_{\;\;\dot\sigma} C^{\dot\sigma}_{\;\;\alpha}-2\alpha'k_1^qk^m_1e_m^{(2)}\xi_{(1)}^\alpha\bar\rho_{q\dot\beta}C^{\dot\beta}_{\;\;\alpha}+e_m^{(2)}\xi_{(1)}^\alpha\bar\rho_{m\dot\sigma}C_{\;\alpha}^{\dot\sigma}\biggl]\non\\
\label{achipsiPS}
\ee
To compare the results involving the fermionic fields with the PS results, we need to first convert the RNS gamma matrix conventions into the PS gamma matrix conventions. This mainly involves setting the charge conjugation matrix to be the Kronecker delta $ \delta^{\alpha}_{\;\;\beta}$ which implies (for details, see e.g., \cite{Hartl:2010ks, Oliver1})
\be
(\gamma^mC)_{\alpha\beta}\rightarrow \gamma^m_{\alpha\beta}\quad,\quad (\bar\gamma_mC)^{\dot\alpha\dot\beta}\rightarrow \gamma_m^{\alpha\beta}\quad,\quad (\gamma^{mnp}C)_{\alpha\beta}\rightarrow (\gamma^{mnp})_{\alpha\beta}
\ee
and so on.

\section{Theta expansion of massless vertex operator}
\label{appen:massless_theta}
Theta expansion of the massless vertex operator is known and has been extensively used in the literature (see, e.g., \cite{mafrathesis}). However, some of our conventions (e.g., equation \eqref{cov_der}) are different from the literature in which the theta expansion of massless vertex operator is used. Below, we derive the results in our convention briefly indicating the steps. We start by recalling the $\mathcal{N}=1$ SYM equations in 10 dimensions \cite{Witten:1985nt}.
\subsection*{$\mathcal{N}=1$ SYM equations in 10 dimensions}
\label{N=1SYM}
\vspace*{.07in}In 10 dimensions, the open string massless states are described by the 10 dimensional $\mathcal{N}=1$ super Yang Mills equations. The field strengths describing the theory are given by
\be
F_{\alpha\beta} = \{\nabla_\alpha,\nabla_\beta\} -2 \gamma^m_{\alpha\beta}\nabla_m\qquad,\qquad
F_{\alpha m} = [\nabla_\alpha,\nabla_m]=-F_{m\alpha}\qquad,\qquad
F_{mn} = [\nabla_m,\nabla_n] \non
\ee
where $ \nabla_m \equiv \p_m+ A_m\ ,\ \nabla_\alpha \equiv D_\alpha+ A_\alpha$ and $D_\alpha$ is defined in \eqref{cov_der}. 

\vspace*{.07in}The 10 dimensional Yang-Mills equations of motion follow from
\be
F_{\alpha\beta} =0 \quad\implies\quad D_\alpha A_\beta +D_\beta A_\alpha = 2\gamma^m_{\alpha\beta} A_m\label{C.11as}
\ee
Using this along with the Bianchi identities, we obtain the following equations at linearized level
\be
F_{ m\alpha} &=& (\gamma_m)_{\alpha\beta}W^\beta \qquad,\qquad
D_{\alpha }W^\beta =-\f{1}{2} (\gamma^{mn})_{\alpha}^{\;\;\beta}F_{mn}\quad\implies\quad D_\alpha W^\alpha=0\non\\[.3cm]
D_{\alpha }F_{mn} &=&\p_m(\gamma_nW)_\alpha-\p_n(\gamma_mW)_\alpha\qquad,\qquad
D_{\alpha }A^{m} =-\gamma^m_{\alpha\beta}W^\beta +\p^m A_\alpha
\non\\[.3cm]
\p_m F^{mn}&=&0 \qquad,\qquad
\gamma^m_{\alpha\beta}\p_m W^\beta =0
\ee
Further, the superfields $A_m, W^\alpha$ and $F_{mn}$ can be expressed as 
\be
A_m &=& \f{1}{16}\gamma_m^{\alpha\beta}D_\alpha A_\beta \qquad,\qquad
F_{mn}= \p_m A_n- \p_n A_m\non\\[.2cm]
W^\alpha &=& -\f{1}{10}(\gamma^m)^{\alpha\beta}(D_\beta A_m -\p_m A_\beta) 
\ee
\subsection*{$\mathcal{N}=1$ SYM equations from pure spinor formalism}
The pure spinor formalism gives the $\mathcal{N}=1$ SYM equations through its BRST equations of motion. To obtain the correct normalization of the superfields, we derive these equations using the BRST equation of motion $QU=\p V$ and match with the equations given above. At the massless level, the unintegrated vertex operator is constructed from the ghost number 1 and conformal weight 0 objects. The most general object with this property has the form $V=\lambda^\alpha A_\alpha$. Similarly, the most general integrated vertex operator has the form 
\be
U= \p\theta^\alpha \tilde A_\alpha +\Pi^m \tilde A_m +d_\alpha \tilde W^\alpha + N^{mn}\tilde F_{mn}
\ee
Using the OPEs given in section \ref{sec:review}, we obtain
\be
Q(\p\theta^\alpha \tilde A_\alpha) &=& -\f{\alpha'}{2}\p\theta^\alpha \lambda^\beta D_\beta \tilde A_\alpha +\f{\alpha'}{2} \p\lambda^\alpha \tilde A_\alpha
\non\\[.2cm]
Q(\Pi^m \tilde A_m) &=& \f{\alpha'}{2}\Pi^m\lambda^\alpha D_\alpha \tilde A_m +\f{\alpha'}{2} \p\theta^\alpha \lambda^\beta \tilde A_m\gamma^m_{\alpha\beta}\non\\[.2cm]
Q(d_\alpha\tilde  W^\alpha) &=& -\f{\alpha'}{2}d_\alpha\lambda^\beta D_\beta \tilde W^\alpha -\f{\alpha'}{2} \Pi^m \lambda^\alpha \tilde W^\beta\gamma^m_{\alpha\beta}+\f{(\alpha')^2}{2}  \p\lambda^\alpha \p_m\tilde W^\beta\gamma^m_{\alpha\beta}\non\\[.2cm]
Q(N^{mn}\tilde F_{mn}) &=& \f{\alpha'}{2}N^{mn}\lambda^\beta D_\beta \tilde F_{mn} -\f{\alpha'}{4} d_\alpha \lambda^\beta \tilde F_{mn}(\gamma^{mn})_{\;\;\beta}^{\alpha}-\f{(\alpha')^2}{8}  \p\lambda^\alpha D_\beta \tilde F_{mn}(\gamma^{mn})_{\;\;\alpha}^{\beta}\non
\ee
We also have,
\be
%\p V &=&\p (\lambda^\alpha A_\alpha) \non\\[.2cm]
\p (\lambda^\alpha A_\alpha)&=&\p\lambda^\alpha A_\alpha +\lambda^\alpha \p A_\alpha\non\\[.2cm]
&=&\p\lambda^\alpha A_\alpha + \p\theta^\beta \lambda^\alpha D_\beta A_\alpha +2 \Pi^m\lambda^\alpha \p_m A_\alpha
\ee
where we have used the pure spinor open string identity 
\be
\p = \p\theta^\alpha D_\alpha +2\Pi^m\p_m
\ee
The BRST equation of motion gives,
\be
0&=&QU-\p V\non\\
&=& \p\theta^\alpha \lambda^\beta \Bigl(-\f{\alpha'}{2} D_\beta \tilde A_\alpha-D_\alpha A_\beta+\f{\alpha'}{2}  \tilde A_m\gamma^m_{\alpha\beta}\Bigl)+\ \Pi^m\lambda^\alpha\Bigl(\f{\alpha'}{2} D_\alpha \tilde A_m-\f{\alpha'}{2}  \tilde W^\beta\gamma^m_{\alpha\beta}-2\p_m  A_\alpha\Bigl) \non\\[.2cm]
&&+\ d_\alpha\lambda^\beta\Bigl(-\f{\alpha'}{2} D_\beta \tilde W^\alpha-\f{\alpha'}{4}  \tilde F_{mn}(\gamma^{mn})_{\;\;\beta}^{\alpha}\Bigl) \ +\ \f{\alpha'}{2}N^{mn}\lambda^\beta D_\beta \tilde F_{mn}  
\non\\[.2cm]
&&+\ \p\lambda^\alpha\Bigl(\f{\alpha'}{2} \tilde A_\alpha+\f{(\alpha')^2}{2}   \p_m\tilde W^\beta\gamma^m_{\alpha\beta}-\f{(\alpha')^2}{8}   D_\beta \tilde F_{mn}(\gamma^{mn})_{\;\;\alpha}^{\beta}-A_\alpha\Bigl)
\ee
To match these equations with the 10 dimensional SYM equations, we rescale the fields as 
\be
\tilde A_\alpha = \f{2}{\alpha'}A_\alpha \quad,\quad \tilde A_m = \f{4}{\alpha'}A_m  \quad,\quad \tilde W^\alpha = -\f{4}{\alpha'}W^\alpha\quad,\quad \tilde F_{mn} = \f{4}{\alpha'}F_{mn}
\ee
After this rescaling, the BRST equation gives
\be
&& D_\beta A_\alpha
+D_\alpha A_\beta=2 A_m\gamma^m_{\alpha\beta}\quad,\quad  D_\alpha A_m= -W^\beta\gamma^m_{\alpha\beta}+\p_m A_\alpha\quad,\quad D_\beta W^\alpha=-\f{1}{2}  F_{mn}(\gamma^{mn})_{\beta}^{\;\;\alpha}\non\\[.3cm]
&&  -\p_mW^\beta\gamma^m_{\alpha\beta}=\f{1}{4}   D_\beta F_{mn}(\gamma^{mn})_{\;\;\alpha}^{\beta}\quad,\quad N^{mn}\lambda^\beta D_\beta F_{mn}  =0\non
\ee
The first 4 equations are precisely satisfied by the 10 dimensional $\mathcal{N}=1$ SYM equations given in appendix \ref{N=1SYM} whereas the last equation is satisfied by the pure spinor constraint. The correctly normalized vertex operators of the massless states in our conventions thus become
\be
V&=&\lambda^\alpha A_\alpha\non\\[.2cm]
U&=&\f{2}{\alpha'} \p\theta^\alpha  A_\alpha + \f{4}{\alpha'}\Pi^m  A_m - \f{4}{\alpha'}d_\alpha  W^\alpha + \f{4}{\alpha'} N^{mn} F_{mn}\non\\[.2cm]
&=&\f{4}{\alpha'} \left(\f{1}{2}\p\theta^\alpha  A_\alpha +\Pi^m  A_m -d_\alpha  W^\alpha + N^{mn} F_{mn}\right)
\ee
\subsection*{Theta expansion of massless superfields}
Finally, we turn to the theta expansion. We shall follow the steps outlined in \cite{Policastro}. We shall need the following equations for doing the theta expansion
\be
&&D_\alpha A_\beta+D_\beta A_\alpha
=2 A_m\gamma^m_{\alpha\beta}\quad,\qquad D_\alpha A_m= -W^\beta\gamma^m_{\alpha\beta}+\p_m A_\alpha \non\\[.2cm]
&&D_\beta W^\alpha=-\f{1}{2}  F_{mn}(\gamma^{mn})_{\beta}^{\;\;\alpha}\quad,\quad F_{mn} =\p_mA_n-\p_nA_m\label{3.3.2}
\ee
Before proceeding to do the theta expansion, we need to fix a gauge. We shall choose the gauge $\theta^\alpha A_\alpha=0$. In this gauge choice, we have
\be
0=D_{\beta}(\theta^\alpha A_\alpha)= A_\beta - \theta^\alpha D_\beta A_\alpha\quad\implies\qquad  A_\beta =\theta^\alpha D_\beta A_\alpha
\ee
Now, multiplying by $\theta^\beta$ in the 1st equation and by $\theta^\alpha$ in the 2nd and 3rd equations of \eqref{3.3.2} and using the above identity along with the gauge choice $\theta^\alpha A_\alpha=0$, we obtain
\be
(1+D)A_\alpha &=& 2A_m (\gamma^m\theta)_\alpha
\quad,\quad 
D A_m = -(\theta\gamma_m W)\quad,\quad
 DW^\alpha =\f{1}{2}F_{mn}(\gamma^{mn}\theta)^\alpha\label{thetac.22}
\ee
where we have defined $D\equiv \theta^\alpha D_\alpha = \theta^\alpha \p_\alpha$.

\vspace*{.07in}We can use the above 3 equations along with the 4th equation of \eqref{3.3.2} to do the theta expansion. If we denote the $\ell^{th}$ order component of the superfield $M$ by $M^{(\ell)}$, then we have
\be
M^{(\ell)} = m_{\alpha_1\cdots \alpha_\ell}\theta^{\alpha_1}\cdots \theta^{\alpha_\ell}\quad\implies \quad DM^{(\ell)}=\ell\ M^{(\ell)}
\ee
Using this, the 3 equations of \eqref{thetac.22} and the 4th equation of \eqref{3.3.2} give the following recursive relations
\be
(1+\ell)A^{(\ell)}_\alpha &=& 2A^{(\ell-1)}_m (\gamma^m\theta)_\alpha\quad\implies\quad A^{(\ell)}_\alpha = \f{2}{1+\ell}A^{(\ell-1)}_m (\gamma^m\theta)_\alpha
\non\\[.3cm]
\ell A^{(\ell)}_m &=& -(\theta\gamma_m W^{(\ell-1)})\quad\implies\quad  A^{(\ell)}_m= -\f{1}{\ell}(\theta\gamma_m W^{(\ell-1)})  \non\\[.3cm]
\ell W^\alpha_{(\ell)} &=&\f{1}{2}F^{(\ell-1)}_{mn}(\gamma^{mn}\theta)^\alpha\quad\implies\quad  W^\alpha_{(\ell)} =\f{1}{2\ell}(\gamma^{mn}\theta)^\alpha\ F^{(\ell-1)}_{mn}\non\\[.3cm]
F^{(\ell)}_{mn}& =&\p_mA^{(\ell)}_n-\p_nA^{(\ell)}_m
\ee
Denoting the theta independent components of the superfields $A_m$ and $W^\alpha$ to be
\be
A_m^{(0)} \equiv a_m\quad,\qquad W^{\alpha}_{(0)}\equiv\chi^\alpha
\ee
the above recursive relations give 
\be
A_\alpha &=& a_m(\gamma^m\theta)_\alpha -\f{2}{3}(\gamma^m\theta)_\alpha (\theta\gamma_m\chi)-\f{1}{8}(\gamma_m\theta)_\alpha (\theta\gamma^{mpq}\theta)f_{pq}-\f{i}{15}(\gamma_m\theta)_\alpha (\theta\gamma_p\chi)(\theta\gamma^{mpq}\theta)k_q+\cdots\non\\[.4cm]
A_m&=& a_m -(\theta\gamma_m\chi) -\f{1}{4}(\theta\gamma_{mnp}\theta)f^{np}\ -\ \f{1}{6}(\theta\gamma_{m}\gamma^{pq}\theta)(\theta\gamma_{[m}\p_{n]}\chi) +\f{1}{48}  (\theta\gamma_{m}\gamma^{rn}\theta)(\theta\gamma_{npq}\theta)\p_rf^{pq}+\cdots\non\\
\ee
where $f_{mn}=\p_m a_n-\p_na_m$ and the plane wave expansion of the gluon and gluino are given by
\be
a_m \ =\ e_m e^{ik\cdot X} \quad,\qquad \chi^\alpha \ =\ \xi^\alpha e^{ik\cdot X}
\ee

\end{document}